\documentclass[aps,pre,amsmath,amsfonts,amssymb,11pt,nofootinbib,showpacs]{revtex4}
\usepackage{color,epsfig,graphics,psfrag,rotating,theorem}

\newtheorem{lemma}{Lemma}

\newtheorem{propo}{Proposition}
\newtheorem{conj}{Conjecture}

\newcommand{\eps}{\varepsilon}

\def\Tree{{\mathbb T}}
\def\T{{\sf T}}

\def\E{\mathbb E}
\def\prob{{\mathbb P}}

\def\ed{\stackrel{\rm d}{=}}

\def\P{{\sf P}}
\def\M{{\mathfrak M}_q}
\def\Q{\widehat{Q}}
\def\QQ{\widetilde{Q}}
\def\W{\widehat{W}}
\def\F{{\sf F}}

\def\pih{\widehat{\pi}}

\def\eo{\overline{\eta}}

\def\ux{\underline{x}}
\def\uX{\underline{X}}
\def\uY{\underline{Y}}
\def\uy{\underline{y}}
\def\de{{\rm d}}
\def\cG{{\cal G}}
\def\cN{{\cal N}}

\def\prooft{\hspace{0.5cm}{\bf Proof:}\hspace{0.1cm}}
\def\endproof{\hfill$\Box$\vspace{0.4cm}}

\begin{document}

\title{Reconstruction on trees and spin glass transition}

\author{Marc M\'ezard}

\affiliation{Laboratoire de Physique Th\'eorique et Mod\`eles Statistiques,
Universit\'e de Paris-Sud, b\^atiment 100, 91405, Orsay Cedex, France}

\author{Andrea Montanari}

\affiliation{Laboratoire de Physique Th\'eorique de l'Ecole Normale
Sup\'erieure,\\
 24 rue Lhomond 75231 Paris Cedex 05, France}


\begin{abstract}
Consider an information source generating a symbol at the 
root of a tree network whose links correspond to noisy
communication channels, and broadcasting it through the network.
We study the problem of reconstructing the transmitted
symbol from the information received at the leaves.
In the large system limit, reconstruction is possible when the 
channel noise is smaller than a threshold. 

We show that this threshold coincides with the dynamical 
(replica symmetry breaking) glass transition for an associated 
statistical physics problem. Motivated by this correspondence,
we derive a variational principle which implies new rigorous bounds 
on the reconstruction threshold. Finally, we apply a standard
numerical procedure used in statistical physics, 
to predict the reconstruction thresholds in various channels. 
In particular, we prove a bound on the
reconstruction problem for  the antiferromagnetic ``Potts'' channels, which
implies, in the noiseless limit, new results on random  proper colorings of 
infinite regular trees.

This  relation to the reconstruction problem also offers  interesting perspective for
putting  on a clean mathematical basis the theory of  glasses on random graphs.
\end{abstract}


\maketitle

\section{Introduction}
\label{sec:intro}

\begin{figure}
\includegraphics[width=0.4\linewidth,angle=-0]{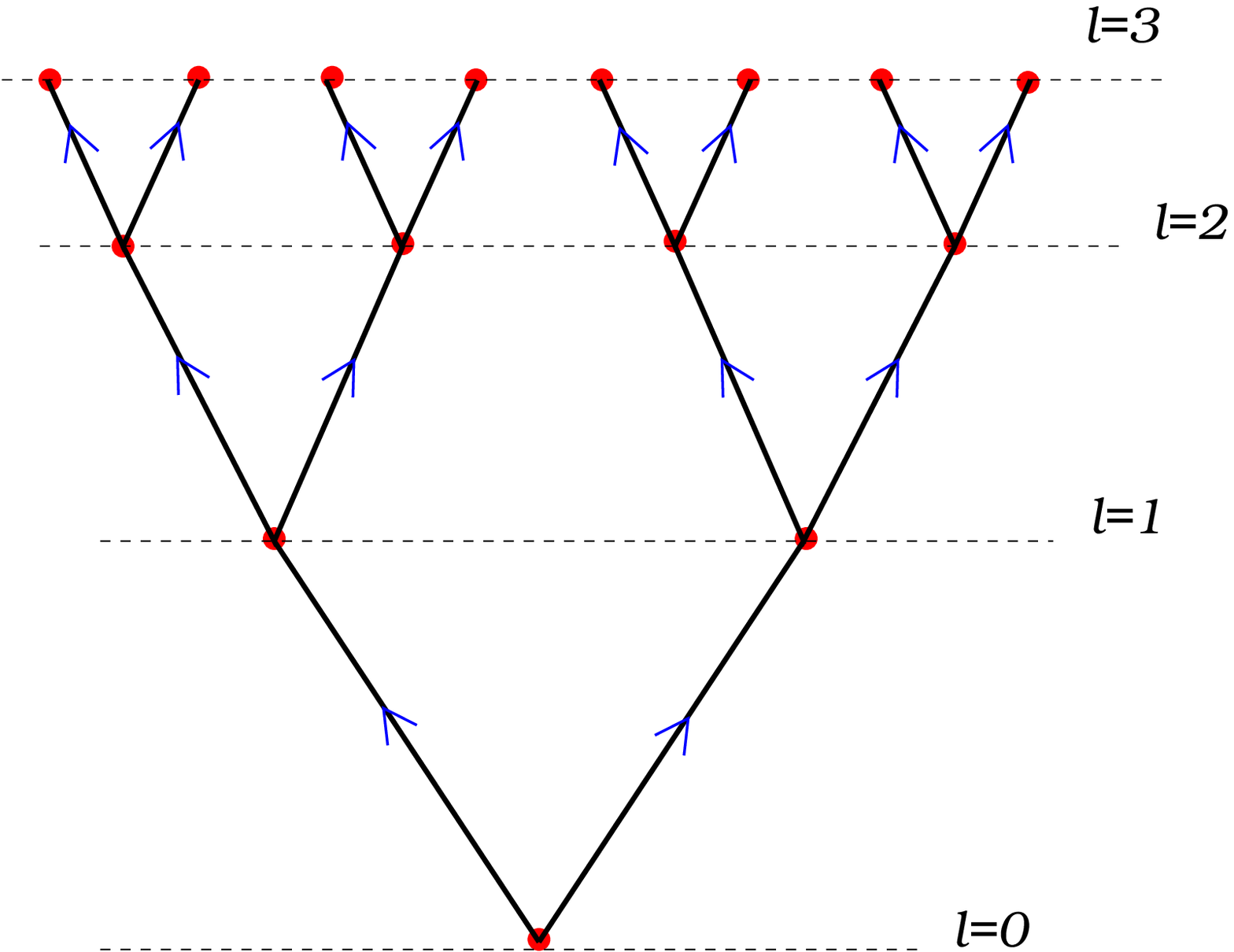}\hspace{1.5 cm}
\includegraphics[width=0.38\linewidth,angle=-0]{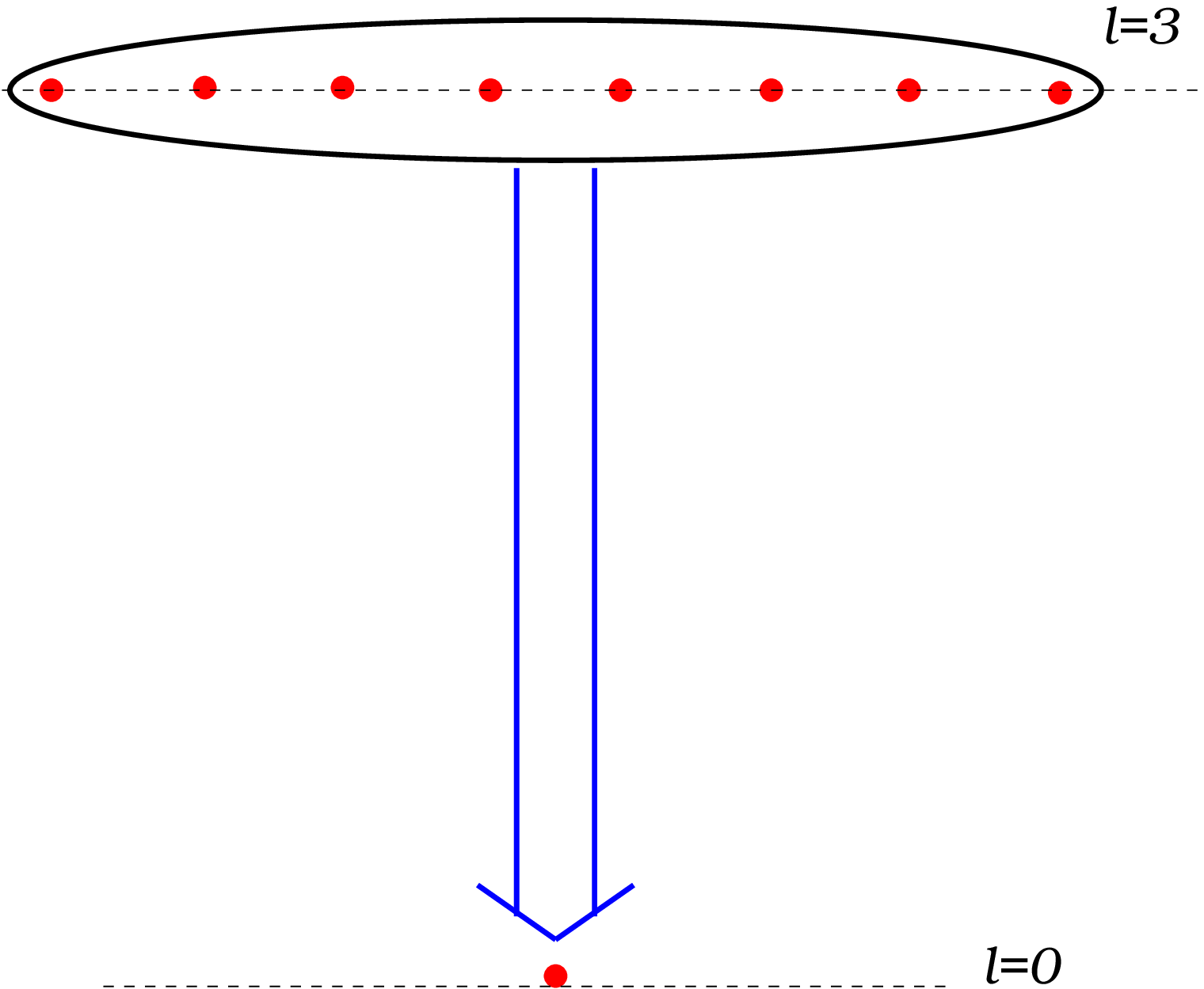}
\caption{Left: Broadcast on a tree. The signal is sent from the root. Each
edge is a noisy communication channel broadcasting upwards. Right: The reconstruction problem asks to find what 
signal was sent from the root, given the signals received on the leaves }
\label{fig:broadcast}
\end{figure}

Consider the following broadcast problem~\cite{Mossel03}.
An information source at the root of a tree network
produces a letter taken from a $q$-ary alphabet
$x\in \{1,\dots,q\}$ (we shall sometimes refer to a letter 
from this alphabet as to a `color'). 
The symbol is propagated along the edges of the tree. 
For simplicity we start with a regular $k$-ary tree $\Tree_k$, 
cf. Fig.~\ref{fig:broadcast},
in which every vertex has exactly $k$ descendants 
(every vertex has degree $k+1$ except the root which
has degree $k$), a more general setting is described in~\cite{EvaKenPerSch} and in Sect.~\ref{se:gene}. 
Each edge of the tree is an instance of the
same noisy communication channel: If the letter $x$ is transmitted 
through the channel, $y\in\{1,\dots,q\}$  is received with
probability $\pi(y|x)$ (with $\pi(y|x)\ge 0$ and $\sum_y\pi(y|x)=1$). 
The problem of reconstruction is the following:  consider all the
symbols received at the vertices of the $\ell^{\rm th}$ generation. Does this
configuration contain a non-vanishing information on the letter transmitted 
by the root, in the large $\ell$ limit?

Beyond its fundamental interest in
probability, this problem is relevant to genetics (propagation of genes
from an ancestor) \cite{MosselSteel05}, 
to statistical physics (models on  Bethe lattices),
and information theory (the problem being equivalent to computing
the information capacity of the tree network)
\cite{CoverThomas}.

An important general bound was obtained by Kesten and Stigum (KS)
\cite{KestenStigum1,KestenStigum2}. 
Consider the matrix $\pi$ with entries $\pi(y|x)$,
$x,y\in\{1,\dots,q\}$ and let $\lambda_2(\pi)$ be its
eigenvalue with the second largest absolute value. Then, if $k 
|\lambda_2(\pi) |^2 >1$, the reconstruction problem is solvable: the leaves
asymptotically contain some information on the letter sent by the root. 
In fact in this case the census of the variables in the $\ell^{\rm th}$ 
generation (the number of leaves which have received each letter) 
contains some information on the root. 
Conversely, if $k | \lambda_2(\pi)|^2 <1$,
the census contains asymptotically no information on the root
 \cite{MosselPeres01}. 
Therefore, the
KS condition $k \vert \lambda_2(\pi)\vert ^2=1$ defines a
threshold for the maximum amount of noise allowing census
reconstruction. For larger noise ( $k \vert \lambda_2(\pi)\vert ^2 <1$)
one may wonder whether reconstruction is possible exploiting the 
whole set of symbols received at the
$\ell^{\rm th}$ generation, through a clever use of the correlations 
between the symbols received on the leaves. The answer depends
on the channel.

In most of this paper we shall focus onto
transition kernels $\pi(\,\cdot\,|\,\cdot\,)$ satisfying
the detailed balance condition (reversible) 
with respect to the uniform distribution $\eo(x) = 1/q$. 
In other words $\pi(y|x) = \pi(x|y)$. With a slight
abuse of notation we shall write  $\pi(y|x) = \pi(y,x)$.
For the problem to be non-trivial, we also assume 
$\pi(\,\cdot\,|\,\cdot\,)$ to be irreducible and aperiodic.
A particularly important example in this family is provided by
$q$-ary symmetric channels (or, borrowing from the statistical mechanics
terminology, `Potts' channels)
\begin{equation}
\pi(y|x) = \left\{\begin{array}{ll} 1-\eps & \;\; \mbox{if $y=x$.} \\
		\eps/(q-1)  & \;\; \mbox{otherwise}\end{array}
\right. \ .\label{eq:PottsChannelDef}
\end{equation}
If $\eps<1-1/q$, $y=x$ is the most likely channel output when the input
is $x$: we shall refer to this case as the `ferromagnetic' Potts channel.
If $\eps>1-1/q$, the opposite happens and we shall speak of
`antiferromagnetic' Potts channel. The particular case 
$\eps=1$ is of special interest, since the broadcast 
process provides a uniformly random proper coloring of the 
$\ell$-generations $k$-ary tree $\Tree_k(\ell)$.

It is intuitively clear that the channel (\ref{eq:PottsChannelDef}) 
`gets worse' as $\eps$ increases from $0$ to $1-1/q$ (ferromagnetic channel) and 
`improves' as $\eps$ goes from $1-1/q$ to $1$ (antiferromagnetic
channel).
A result by Mossel~\cite{Mossel01} implies that there exist 
a ferromagnetic and an antiferromagnetic threshold,
respectively $\eps^+_{\rm r}(k,q)\in[0,1-1/q]$ and 
$\eps^-_{\rm r}(k,q)\in[1-1/q,1]$,
such that the reconstruction problem is solvable when
$\eps \in [0,\eps^+_{\rm r}[\; \cup\;  ]\eps^-_{\rm r},1]$ and insolvable
if $\eps \in \; ]\eps^+_{\rm r},\eps^-_{\rm r}[$. 
Hereafter we shall drop the $\pm$ superscripts whenever they are clear from the context.

The KS condition $k|\lambda_2(\pi)|^2>1$ is satisfied 
(and the problem is census-solvable) for  the channel 
(\ref{eq:PottsChannelDef})  if and only if 
$\eps\in [0,\eps_{\rm KS}^+(k,q)[\;  \cup\;  ]\eps_{\rm KS}^-(k,q),1]$, 
where: 
\begin{eqnarray}
\eps_{\rm KS}^{\pm}(k,q) &=& \frac{q-1}{q} \; \left(1\mp
\frac{1}{\sqrt{k}}\right)\, .
\end{eqnarray}
Notice that the above formula yields $\eps_{\rm KS}^{-}(k,q)>1$ for some pairs
of $(k,q)$. In fact, for the antiferromagnetic channel, the 
census-reconstruction problem (as well as the general reconstruction problem)
is not necessarily solvable for $\eps=1$.

It is known \cite{BleherRuizZagrebnov} that, for $q=2$ (the ``binary
symmetric'' channel, also known as the ``symmetric Ising'' case), the
reconstruction threshold is equal to the KS one: 
$\eps_{\rm r}(k,2)=\eps_{\rm KS}(k,2)$ (for $q=2$ the ferromagnetic
and antiferromagnetic cases are equivalent via the mapping
$\eps\mapsto 1-\eps$). In general, the KS bound implies 
$\eps^{+}_{\rm KS}(k,q)\ge \eps_{\rm r}^{+}(k,q)$, and
$\eps^{-}_{\rm r}(k,q)\ge \eps_{\rm KS}^{-}(k,q)$. 
Furthermore, in \cite{Mossel01} it was shown that, for all $k$, when
$q$ is large enough, $\eps^{+}_{\rm KS}(k,q)> \eps_{\rm r}^{+}(k,q)$
strictly: reconstruction is possible at noise levels where
census reconstruction does not work. 
However, several fundamental 
questions remain open even for simple Potts channels:
Is there any pair $(k,q)$, with $q>2$, such that 
$\eps_{\rm r}(k,q)=\eps_{\rm KS}(k,q)$? How to distinguish systematically 
between $\eps_{\rm r}(k,q)$ and $\eps_{\rm KS}(k,q)$? How to determine
$\eps_{\rm r}(k,q)$ accurately when it does not coincide with 
$\eps_{\rm KS}(k,q)$? We shall address these issues in the following. 

The reconstruction problem is intimately related to statistical physics. 
Consider a model of Potts spins $y_i \in \{1,\dots,q\}$, 
on a finite rooted tree with $\ell$ generations, to be denoted by 
$\Tree_k(\ell)$.  
Suppose that the energy of a configuration $\uy^{\ell}\equiv \{y_i :
i\in \Tree_k(\ell)\}$ is given by:
\begin{equation}
E(\uy^{\ell})= - J \!\!\sum_{(i,j)\in \Tree_k(\ell)}\delta_{y_i, y_j} \ ,
\label{eq:PottsEnergy}
\end{equation}
where $(i,j)$ denotes pairs of spins connected by an edge of the
tree. Let $\uY^{\ell}$ be the random configuration produced by the broadcast
process with channel (\ref{eq:PottsChannelDef}) {\it  up to} generation 
$\ell$, 
when the transmitted symbol is uniformly random in $\{1,\dots,q\}$.
Then
\begin{eqnarray}
\prob\left\{\uY^{\ell}=\uy^{\ell}\right\}
=\frac{1}{Z}\, \exp\{-\beta E(\uy^{\ell})\}
\end{eqnarray}
provided we make the identification
\begin{equation}
e^{-\beta J}=\frac{\eps}{(q-1) (1-\eps)}\ .
\end{equation}
In other words, the broadcast process allows to construct one 
particular Gibbs measure (state) associated to the energy 
function (\ref{eq:PottsEnergy}): in the
statistical physics terminology this is the free-boundary measure.
In general this is not the unique Gibbs measure for 
this energy function. For instance, if $\eps< \frac{q-1}{q} \;
\left(1-\frac{1}{k}\right)$, one can construct $q$ `ferromagnetic' states
as well. Even if more than one Gibbs state exists, the free-boundary 
state can be extremal (or `pure'). It turns out that
 the reconstruction problem is 
solvable if and only if the Gibbs state with free boundary conditions
is not extremal.

Given the strong connection between extremality of Gibbs states and 
spatial decay of correlations \cite{Georgii}, the last remark is not surprising.
What is more surprising (and constitutes the main theme of this paper)
is the relation of the reconstructibility with the existence of 
a dynamical glass phase. In recent years, an ongoing effort has been 
devoted to the study of glassy models on sparse random graphs.
These are graphs which contain cycles but locally `look like' a tree
(e.g. uniformly random graphs with given degree).
One of the most widespread features of these models, is the
occurrence of glass phases in which the Boltzmann measure gets split
into an exponential number of `lumps' (also referred to as clusters or 
pure states). This phenomenon is usually studied by solving
some `one-step replica symmetry breaking' (1RSB) distributional equations.
In the following we show that these equations, as well as the criterion 
used to detect glass phases, do indeed coincide 
with the solvability of an appropriate reconstruction problem. 

In spin glass theory, one can encounter
two types of transitions to a glass phase. In the first 
case the transition is continuous in a properly defined order parameter.
In spin glass jargon this leads to a 
phase with `full replica symmetry breaking' (FRSB).
In the second it is discontinuous, leading to 1RSB.
Both situations occur in the reconstruction problem, depending on the alphabet
and the channel. In the continuous case, the phase transition
location is given by a local instability which coincides with the 
KS threshold, and one has $\eps_{\rm r}(k,q)=\eps_{\rm KS}(k,q)$.
This happens, for instance, when $q=2$.
In the opposite case, the `dynamical' glass transition is discontinuous and its
location (which still coincides with the reconstruction threshold)
is distinct from the KS one. In the ferromagnetic Potts model one has, 
for instance,  $\eps_r(k,q)>\eps_{KS}(k,q)$ at large enough $q$.

The coincidence of the reconstruction threshold with the dynamical 
glass transition, apart from being interesting in itself, allows us
to adapt several techniques developed within the theory of spin glasses
in order to study the reconstruction problem.
On the one hand, importing a numerical procedure currently used in this field, we 
determine the threshold for several pairs $k,q$. These results 
lead us to conjecture that $\eps_{\rm r}(k,q)=\eps_{\rm KS}(k,q)$,
for $k$ not too large and $q\le 4$ (in the ferromagnetic case) or $q\le 3$ 
(in the antiferromagnetic case).

Furthermore, we derive a variational principle for the reconstruction problem. 
In the antiferromagnetic case, this implies
a rigorous bound on the reconstruction threshold, which allows to confirm 
the strict inequality $\eps^{-}_{\rm r}(k,q)<\eps^{-}_{\rm KS}(k,q)$ 
in most of the cases in which this was found to be the case numerically.
Although we conjecture such a bound to hold in much greater generality,
we weren't able to prove it, and we leave it as a conjecture.

The paper is organized as follows. In Section~\ref{sec:Recursion}
we define the main objects studied in the paper and prove the coincidence 
between reconstruction and dynamical glass transition.
In Section \ref{se:varprinc} we state our variational principle
and prove that it provides a rigorous bound for a class of 
kernels $\pi(\,\cdot\,|\,\cdot\,)$ including the antiferromagnetic model.
In Secs. \ref{sec:Ferro} and \ref{sec:AntiFerro} we apply this principle 
as well as a numerical procedure to the determination of thresholds for
the Potts channel, respectively in the ferromagnetic and antiferromagnetic case. 
Section \ref{se:spinglass} discusses the physical meaning of
the relation between reconstruction and glass transitions. 
Section \ref{se:gene} explains how our methods (and the glass - reconstruction correpondance)
 can be generalized to a broad
category of broadcast and reconstruction problems on trees, going much beyond the Potts channel.
We conclude in Section \ref{sec:Conclusion} by summarizing a few 
conjectures and pointing out some interesting open problems.
%
%
\section{Distributional recursion}
\label{sec:Recursion}

\subsection{Definitions}

We denote by $V$ and $E$ the vertex and edge sets of the infinite 
$k$-ary tree $\Tree_k$, by  $0$ its root and by $V_{\ell}$ the set of 
generation-$\ell$ vertices ($|V_{\ell}|= k^{\ell}$). The broadcast process 
generates a random color
configuration $\uX\equiv\{ X_i\; :\;i\in V\}$ with $X_i\in \{1,\dots, q\}$.
The root color $X_0 \in \{1,\dots, q\}$, which we also call the transmitted color,  is uniformly random. 
Then, given the values
of $X$ up to the $\ell$-th generation, the values at the $(\ell+1)$-th
generation are conditionally independent. If a vertex in the $\ell$-th
generation has color $y$, the probability that a vertex connected to it in the
$(\ell+1)$-th generation has color $z$ is $\pi(z|y)$.

We shall denote by $\uX_{\ell}$ the configurations of
colors at the $\ell^{\rm th}$ generation, and by 
$\uY^{\ell}$ the configuration {\em up to} the $\ell^{\rm th}$ generation
(i.e. $\uY^{\ell} = \{\uX_0,\uX_1,\dots,\uX_{\ell}\}$). 
The probability distribution of $\uX_{\ell}$, conditioned to the choice
$X_0=x$ of the root color will be denoted by 
$B^{(\ell)}_x(\ux_\ell)\equiv \prob\{\uX_{\ell} = \ux_{\ell}|X_0=x\}$.

Suppose now that the configuration of colors at the $\ell$-th generation,
$\ux_{\ell}$, is given. We denote by $\eta_{\ell}(y)$
the probability that the root had sent the color $y$, given $\uX_{\ell}$:
\begin{eqnarray}
\eta_{\ell}(y) = \prob[X_0 = y\, |\uX_\ell=\ux_{\ell}]\, .\label{eq:DefEta}
\end{eqnarray}
$\eta_{\ell}(\cdot)$ is a probability distribution over $\{1,\dots, q\}$ (i.e.
$\eta_{\ell}(y)\ge 0$ and $\sum_y\eta_{\ell}(y) = 1$). We shall denote the
space of such distributions as $\M$. 
In order to emphasize the dependency of $\eta_{\ell}$ upon the configuration
received in shell $\ell$, $\ux_{\ell}$,
we shall sometimes write  $\eta_{\ell}(y) =\eta_{\ux_\ell}(y)  $.
It is easy to realize that, given the colors received
at the $\ell^{\rm th}$ generation, $\eta_{\ell}(\cdot)$ constitutes 
a sufficient statistics for the root color $x$. In other terms,
given $\uX_{\ell}=\ux_\ell$, there is no loss of information in computing
$\eta_{\ux_\ell}(\cdot)$ and then guessing $X_0$ from $\eta_{\ux_\ell}(\cdot)$.

Since $\uX_\ell$ is chosen randomly 
according to the broadcast process, $\eta_{\ell}(\cdot)$ is a
random probability distribution, i.e. a random point in $\M$. We
denote by $Q^{(\ell)}_{x}(\eta)$ its 
distribution\footnote{Notice that we adopt here the standard physicists 
convention: we carelessly denote probability distributions by their densities 
even if such densities do not exist. We shall also denote by
$\int \!f(\eta)\, \de Q^{(\ell)}_x(\eta)$ the expectation with respect to 
such a distribution. The fussy reader can easily translate
all the formulae below in the standard probability language.} conditional to
the broadcast being started from $X_0=x$, and call it the `distribution at the root'.
Hereafter a distribution $Q$ over $\M$ will be said {\em trivial} if it is a
singleton on the uniform measure $\eo$ (defined by $\forall x:\ \eo(x) = 1/q$). 
Clearly the reconstruction problem is solvable if and only if the large 
$\ell$ limit of  $Q_{x}^{(\ell)}$ is non trivial. 

There are several ways of characterizing quantitatively the large
$\ell$ behavior of $Q_{x}^{(\ell)}$. We shall consider below two parameters
$I_{\ell} \equiv I(X_0;\uX_\ell)$ and 
$\Psi_{\ell}$, which are defined by:
\begin{eqnarray}
I_{\ell}  =  \frac{1}{q}\sum_x\int\log_2\frac{\eta(x)}{\eo(x)}\; 
\de Q_x^{(\ell)}(\eta)\, ,\;\;\;\;
\Psi_{\ell} = \frac{1}{q}\sum_x\int [\eta(x)-\eo(x)]\;\de 
Q_x^{(\ell)}(\eta)\, .\
\end{eqnarray}

 $I_{\ell}$ gives the number of information bits that can be transmitted reliably 
per network use. $\Psi_{\ell}$ is the probability that the reconstruction
is successful when the receiver guesses color $y$ with probability
$\eta_{\ell}(y)$, minus the the same probability when the receiver 
guesses uniformly.
These are 
non-negative quantities and can be shown to be non-increasing 
functions of $\ell$. We furthermore let $I_{\infty}\equiv\lim_{\ell\to\infty} I_{\ell}$,
and $\Psi_{\infty}\equiv\lim_{\ell\to\infty} \Psi_{\ell}$.

The tree reconstruction problem can be rephrased by saying that 
the problem is solvable if and only if $I_{\infty}>0$ (or, equivalently, 
$\Psi_{\infty}>0$). For instance, for the ferromagnetic
Potts channel, the threshold 
$\eps^+_{\rm r}(k,q)$  is the
supremum of the values of $\eps$ such that $I_{\infty}>0$.
%
%
\subsection{Merging rooted trees}
\label{se:merging}

\begin{figure}
\includegraphics[width=0.3\linewidth,angle=-0]{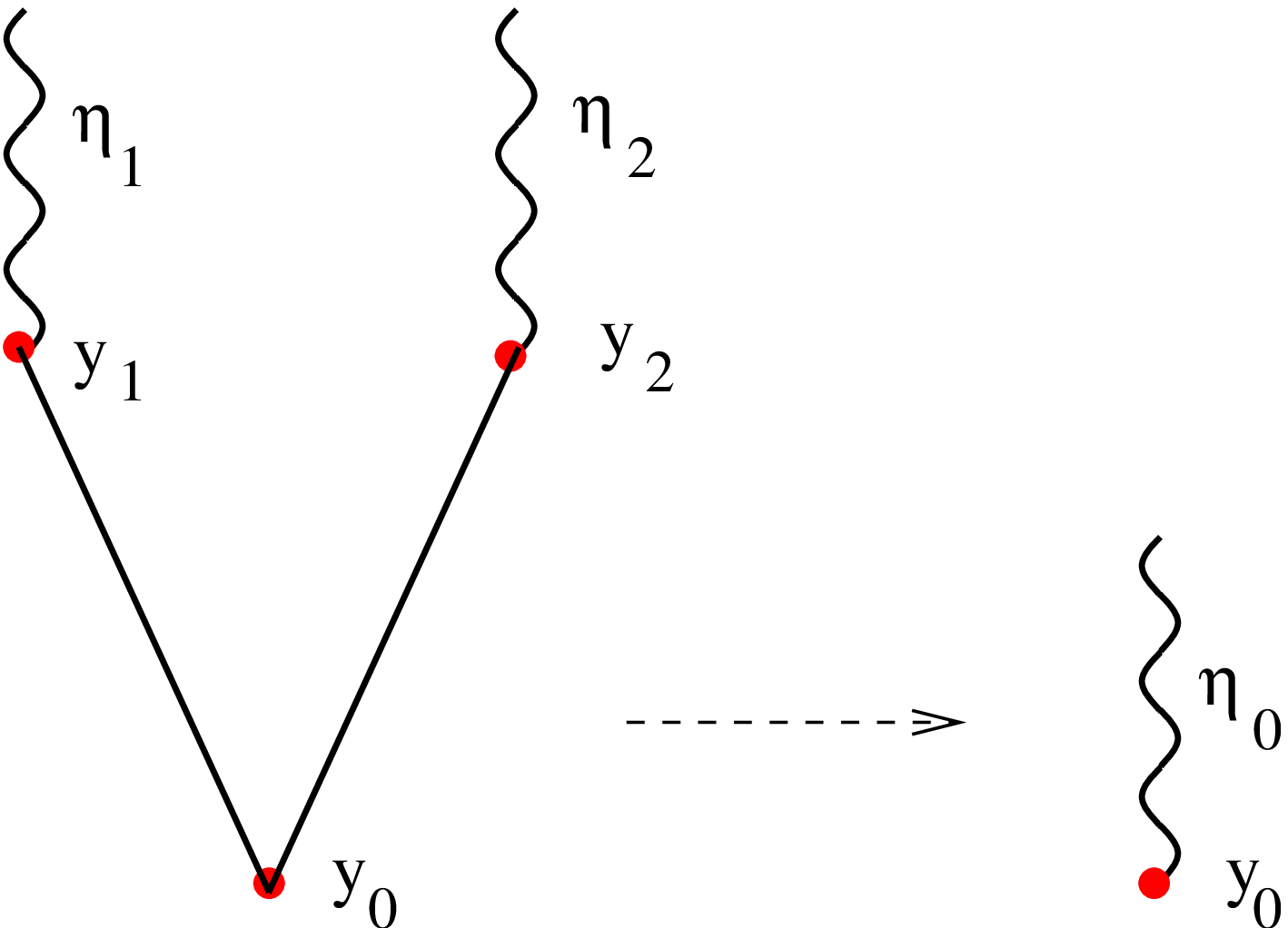}\hspace{1.8 cm}
\includegraphics[width=0.5\linewidth,angle=-0]{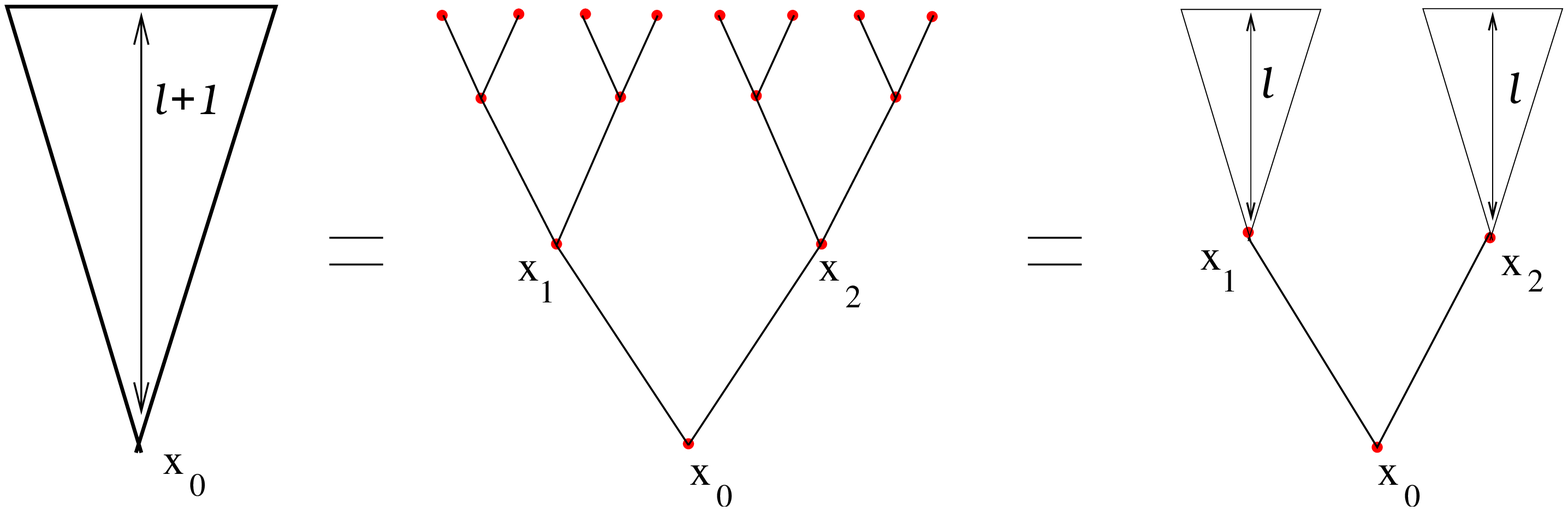}
\caption{Left: a  pictorial representation of the mapping $\eta =
\F(\eta_1,\dots,\eta_k)$ defined in (\ref{eq:Site}). Here $k=2$,
a straight  line corresponds to the channel $\pi$, a wiggly line arriving on
a vertex $y_j$ corresponds to a weight $\eta_j(y_j)$. Right: a  pictorial representation of the
 recursion (\ref{eq:Iteration}). A triangle of depth $r$ rooted on variable $x_j$ denotes $Q_{x_j}^{(r)}(\eta_r)$}
\label{fig:iter}
\end{figure}

How does one compute the distribution $\eta(y)$ on the root, given a boundary
$\uX_{\ell}=\ux_{\ell}$? 
Using the tree-structure, this can be done iteratively by a
dynamical programming procedure starting from the leaves. Suppose that at some
point in this iteration we have determined the probability distributions
$\eta_1(\,\cdot\,),\dots,\eta_k(\,\cdot\,)$ of the $k$ vertices in the 
tree which lie above a given vertex (see Fig.~\ref{fig:iter}, left). 
Then the probability $\eta(y)$ that this vertex had color $y$ 
during the broadcast is given by:
\begin{eqnarray}
\eta(y) = \frac{1}{z(\{\eta_i\})}\, 
\prod_{i=1}^k\left(\sum_{y_i=1}^q\pi(y_i|y)\, \eta_i(y_i)\right)\, ,\;\;
\;\;\;\;\;\; z(\{\eta_i\}) \equiv \sum_{y=1}^q \prod_{i=1}^k\left(\sum_{y_i}\pi(y_i|y)\, \eta_i(y_i)\right)\, .
\label{eq:Site}
\end{eqnarray}
This equation defines a mapping between distributions in $\M$: given $k$
distributions $\eta_1,\dots,\eta_k $, one generates a new one $\eta =
\F(\eta_1,\dots,\eta_k)$. Iterating this mapping
downwards from the leaves down to the root, one can derive the the conditional
distribution of the transmitted symbol.

Equation (\ref{eq:Site}) naturally induces a recursion equation for the 
distribution $Q_{x}^{(\ell)}$.
Consider the reconstruction of the root in a rooted tree with $\ell+1$
generations (see Fig.~\ref{fig:iter}, right). This graph is formed by $k$ subtrees rooted in the vertices
$1,\dots,k$, which are all joined to the root $0$.
Each of these subtrees gives an instance of the reconstruction with $\ell$
generations. 
Therefore:
\begin{eqnarray}
Q_{x}^{(\ell+1)} (\eta)= \sum_{x_1\dots x_k}\prod_{i=1}^k\pi(x_i|x)
\int \; \delta\left[\eta- \F(\eta_1,\dots,\eta_k)\right]\;
\prod_{i=1}^k \de Q^{(\ell)}_{x_i}(\eta_i)
\, ,\label{eq:Iteration}
\end{eqnarray}
where $\delta[\cdots]$ represents a Dirac delta function on $\M$. 
In words, in order to generate $\eta(\,\cdot\, )$ with distribution 
$Q_{x}^{(\ell+1)}$, one can proceed as follows.
First draw $k$ independent colors $x_1,\dots,x_k$
from the distribution $\pi(\,\cdot\,|x)$.
Then, draw $\eta_1,\dots,\eta_k$ independently with distribution, respectively,
 $Q_{x_1}^{(\ell)},\dots,Q_{x_k}^{(\ell)}$. Finally,
let $\eta = \F(\eta_1,\dots,\eta_k)$.

The initial condition is 
\begin{eqnarray}
Q^{(0)}_x(\eta) = \delta\left[\eta-\delta_x\right]\, ,\label{eq:Initial}
\end{eqnarray}
where $\delta_x$ is the distribution in $\M$ which has weight unity on color
$x$ (it is given by $\delta_x(y)=1$ if $y=x$, and $\delta_x(y)=0$ otherwise).
The equations (\ref{eq:Iteration}) and (\ref{eq:Initial}) fully characterize 
the distributions $Q^{(\ell)}_x$. The whole reconstruction problem amounts to
understanding the large $\ell$ properties of these recursions.

%
%
\subsection{Unconditional distribution and symmetry properties}

While $Q^{(\ell)}_x$ gives the distribution of $\eta_{\ell}(\, \cdot\, )$
(defined in Eq.~(\ref{eq:DefEta})) {\em conditional} on  the 
transmitted color being equal to $x$, it is equally interesting to consider the
{\em unconditional} distribution. We will denote it by  $\Q^{(\ell)}$.
Bayes theorem implies the following relation between 
$Q^{(\ell)}_x$ and  $\Q^{(\ell)}$
\begin{eqnarray}
Q^{(\ell)}_{x}(\eta) = q\, \eta(x)\, \Q^{(\ell)}(\eta)\, .
\label{eq:SymmetricRepresentation}
\end{eqnarray}
This is in fact a rephrasing of the identity
\begin{eqnarray}
\prob\{ \eta_{\uX_\ell}=\eta |X_0=x\} = 
\frac{\prob\{X_0=x|\eta_{\uX_\ell}=\eta\}
\prob\{\eta_{\uX_\ell}=\eta\}}{\prob\{X_0=x\}}\, .
\end{eqnarray}

An alternative (analytic) proof can be obtained writing $Q^{(\ell)}_x$
in terms of $B^{(\ell)}_x(\ux_\ell)$, the probability that the output of the 
broadcast process at generation $\ell$ is $\ux_{\ell}$, given that the
transmitted color is $x$:
\begin{eqnarray}
Q^{(\ell)}_x(\eta) = \sum_{\ux_{\ell}} B^{(\ell)}_x(\ux_\ell)\;\; \delta\!\left[
\eta(\cdot)-
\frac{B^{(\ell)}_\cdot(\ux_\ell)}{\sum_z B^{(\ell)}_z(\ux_\ell)}\right]
\, .\label{RepresentationSymm}
\end{eqnarray}
It is then easy to show that, if $\lambda_0+\lambda_1+\dots +\lambda_{q-1}=1$, 
then the expectation value
\begin{eqnarray}
\int  \;  \frac{\eta(0)^{\lambda_0}\cdots \eta(q-1)^{\lambda_{q-1}}}
{\eta(x)} \; \de Q^{(\ell)}_x(\eta) \, 
\end{eqnarray}
does not depend upon $x$. This in turns imply that $Q_x^{(\ell)}$
can be written in the form (\ref{eq:SymmetricRepresentation})
where $\Q^{(\ell)}$ is a distribution which does not depend on $x$ 
(the normalization can be found by summing over $x$).

If the channel  is symmetric with respect to permutations
of the colors (as is the case for Potts channels), the distributions
$Q_x^{(\ell)}$ and $Q^{(\ell)}$ inherit the same symmetry.
More precisely, given a permutation acting on the colors 
$\sigma\in S_q$, and a distribution $\eta\in\M$, let 
$\eta^{\sigma}$ be the permuted distribution defined by
$\eta^{\sigma}(x) \equiv \eta(\sigma(x))$. Then, for any permutation $\sigma$,
$Q^{(\ell)}_x(\eta)=Q^{(\ell)}_{\sigma(x)}(\eta^{\sigma})$, and 
\begin{eqnarray}
\Q^{(\ell)}(\eta^{\sigma}) =  \Q^{(\ell)}(\eta)\, . \label{eq:Symmetry}
\end{eqnarray}
 A distribution satisfying
 condition  (\ref{eq:Symmetry}) will be called  `symmetric'.

Let us finally notice that the parameters introduced in 
Sec.~\ref{sec:Recursion} to measure the amount of information on the transmitted color 
available at the $\ell^{\rm th}$ generation  can be 
expressed in terms of the distribution $\Q^{(\ell)}$
\begin{eqnarray}
I_{\ell}  =  \int D(\eta||\eo)\; \de \Q^{(\ell)}(\eta)\, ,\;\;\;\;
\Psi_{\ell} = \sum_x\int [\eta(x)-\eo(x)]^2\;\de 
\Q^{(\ell)}(\eta)\, .
\end{eqnarray}
Here we use the standard notation for the Kullback-Leibler distance
\cite{CoverThomas} $D(\eta||\eo) \equiv \sum_x \eta(x)\log_2[\eta(x)/\eo(x)]$.
In deriving the second of these expressions, we used the fact that
$\int \eta(x)\, \de\Q^{(\ell)}(\eta) = \eo(x)=\frac{1}{q}$ which follows
from (\ref{eq:SymmetricRepresentation}).
%
%
\subsection{Recursion for the unconditional distribution and 
spin glass correspondence}

The recursion relation (\ref{eq:Iteration}) on $Q^{(\ell)}_x $ 
 implies the following recursion for the unconditional 
distribution:
\begin{eqnarray}
\Q^{(\ell+1)} (\eta)= q^{k-1}\;
\int \;
z(\{\eta_i\})\;\; \delta\left[\eta- \F(\eta_1,\dots,\eta_k)\right]\;
\prod_{i=1}^k \de \Q^{(\ell)} (\eta_i)
\, ,\label{eq:1RSB}
\end{eqnarray}
where $z(\{\eta_i\} )$ is defined as in Eq.~(\ref{eq:Site}).
 The initial  condition (\ref{eq:Initial}) converts into 
$\Q^{(0)}(\eta)=\frac{1}{q} 
\sum_{y=1}^q \delta\left[\eta(\,\cdot\,)-\delta_y(\,\cdot\,)\right]$. 
It is also interesting to study the fixed points of this recursion, 
i.e. the distributions $\Q^{*}$ satisfying:
\begin{eqnarray}
\Q^{*} (\eta)= q^{k-1}\;
\int \;
z(\{\eta_i\})\;\; \delta\left[\eta- \F(\eta_1,\dots,\eta_k)\right]\;
\prod_{i=1}^k \de \Q^{*} (\eta_i)
\, ,\label{eq:1RSB_fixed}
\end{eqnarray}
Notice that any solution of this equation has necessarily
expectation $\int\,\eta(x)\,\de\Q^*(\eta) = \eo(x)$ (this is proved by
taking expectation on both sides). Any probability distribution
over $\M$ satisfying this condition will be hereafter said to be
`consistent'.

The distributional equation (\ref{eq:1RSB_fixed}) is well known in
spin glass theory and usually referred to as `1RSB equation with Parisi parameter $m=1$'
(in the general 1RSB scheme  the factor $z(\{\eta_i\})$ is raised to a  power $m \in [0,1]$).
It is used to determine whether an associated statistical mechanics model
is in a glass phase. We shall return to the definition of the 
associated model in Sec.~\ref{se:spinglass}. For the time being,
we shall adopt the usual physicists criterion as a {\em definition}:
We will say that the statistical mechanics model associated
to the reconstruction problem (characterized by a degree/kernel pair  $k,\pi$) {\em admits a glass phase}
if and only if Eq.~(\ref{eq:1RSB_fixed}) has a non-trivial solution.

When considering a continuous family of kernels $\pi(\,\cdot\,|\, \cdot\,)$,
parametrized by a noise level $\eps$,
the value of $\eps$  where a non-trivial solution appears is 
called a dynamical glass transition.
The result below implies that this coincides indeed with the 
reconstruction threshold (i.e. with the extremality threshold for the free
boundary Gibbs measure on the infinite tree). 
\begin{propo}\label{propo:Reco}
The statistical mechanics model associated
with the degree/kernel pair  $k,\pi$ {\em admits a glass phase}, if 
and only if the corresponding reconstruction problem is solvable.
\end{propo}
\prooft
As  noticed for instance in \cite{Winkler03}, the sequence of 
random variables $\eta_{\ell}(\, \cdot\, )$ (not conditioned on the root 
color), converges almost surely to a limit $\eta_{\infty}(\, \cdot\, )$.
As a consequence, the sequence of distributions $\Q^{(\ell)}$ converges
weakly to the distribution  $\Q^{(\infty)}$ of $\eta_{\infty}(\,\cdot\,)$.
By taking the limit of Eq.~(\ref{eq:1RSB}) (and noticing that 
$\F(\eta_1,\cdots,\eta_k)$ and $z(\{\eta_i\})$ are continuous 
and bounded) we find that $\Q^{(\infty)}$ must satisfy the fixed
point condition (\ref{eq:1RSB_fixed}).
If the reconstruction problem is solvable, then $\Q^{(\infty)}$ is
non-trivial and therefore, according to our definition,
the pair  $k,\pi$ {\em admits a glass phase}.

Conversely\footnote{The idea of the converse is due to James Martin
who kindly agreed to let us publish it here.}, let $\Q^*$ 
be a non-trivial solution of (\ref{eq:1RSB_fixed}). Following
(\ref{eq:SymmetricRepresentation}), define the distribution 
$Q^*_x(\eta) = q\, \eta(x) \Q^*(\eta)$. Because of the above calculations,
the $q$ distributions  $Q^*_x$, $x\in\{1,\dots,q\}$
 are a fixed point of the recursion (\ref{eq:Iteration}). 
We will now show that they 
can be used to reconstruct the transmitted color from the 
output at generation $\ell$, with probability of success independent 
of $\ell$ and strictly larger than $1/q$.

The reconstruction procedure goes as follows.
Suppose that the broadcast has generated the values $X_i=x_i$ for $i\in
V_\ell$. For each vertex $i$, generate $\eta_i$ from the
distribution $Q^*_{x_i}$. Consider now a vertex $a \in
V_{\ell-1}$,connected to 
$a_1,\dots,a_k$ in $V_\ell$. 
Compute $\eta_a= \F(\eta_{a_1},\dots,\eta_{a_k})$, where $\F(\cdots)$ is
defined as in Eq.~(\ref{eq:Site}). Proceeding downwards from the leaves
to the root, this allows to construct $\eta_0$. At this point
the transmitted symbol can be guessed, for instance, by choosing 
$X_0=y$ with probability proportional to $\eta_0(y)$.

We claim that for each vertex $j\in \Tree_k(\ell)$, and conditional to the broadcast having produced
$X_j=x_j$, the  $\eta_j(\,\cdot\,)$ provided by the above procedure is 
distributed according to $Q^*_{x_j}$. This in particular implies
that the probability of guessing correctly the root color is
\begin{eqnarray}
\frac{1}{q}\sum_x\int \eta(x)\;  \de Q_x^*(\eta) =
\sum_x\int \eta(x)^2\;  \de \Q^*(\eta) >\frac{1}{q}\, .
\end{eqnarray}
The claim is proved by induction starting from the leaves 
and proceeding downwards to the root. It is true by construction for
the vertices of the last generation. Assume it to be true
up to generation $r$ and consider a site $a$ in generation
$r-1$ connected to  $a_1,\dots,a_k$ in $V_r$, under the condition
$X_a=x_a$. It is clear that the distribution  of $\eta_a$ is obtained  
through the recursion (\ref{eq:Iteration}) (with $Q^{(\ell)}_{x_i}$
replaced by $Q^{*}_{x_{a_i}}$), and since $Q^{*}_{x_{a_i}}$ is a fixed
point of this recursion, this proves the claim.
\endproof
%
%
\section{Variational principle} 
\label{se:varprinc}
%
%
\subsection{The general principle}

Here we establish a variational principle from which the fixed point equation
(\ref{eq:1RSB_fixed}) for the distribution at the root
can be deduced. We shall not try
to explain here  its physical origin, which is related
to spin glass theory \cite{MP_Bethe}, but just discuss the relation
with the reconstruction problem. Throughout this
section we use the notation $\pi(x|y) = \pi(y|x) \equiv
\pi(x,y)$. Given a distribution $\Q$ over $\M$, we define its {\em complexity}
as 
\begin{eqnarray}
\Sigma(\Q) = -\frac{k+1}{2}\int \W_{\rm e}(\eta_1,\eta_2)\;\; 
\de\Q(\eta_1)\,\de\Q(\eta_2)+ \int   \W_{\rm v}(\eta_1,\dots,\eta_{k+1})\;\;
\prod_{i=1}^{k+1}\de\Q(\eta_i)\, ,
\end{eqnarray}
where
\begin{eqnarray}
\W_{\rm e} &\equiv & - \left[\frac{\sum_{x_1,x_2}\eta_1(x_1)\eta_2(x_2)\pi(x_1,x_2) }
{\sum_{x_1,x_2} \eo(x_1)\eo(x_2)\pi(x_1,x_2)}\right] 
\log\left[\frac{\sum_{x_1,x_2}\eta(x_1)\eta(x_2)\pi(x_1,x_2)}
{\sum_{x_1,x_2}\eo(x_1)\eo(x_2)\pi(x_1,x_2)}\right] \, ,\\
\W_{\rm v} &\equiv & - \left[\frac{\sum_x\prod_i
\sum_{x_i}\eta_i(x_i)\pi(x,x_i)}
{\sum_x\prod_i\sum_{x_i}\eo(x_i)\pi(x,x_i)}\right] 
\log\left[\frac{\sum_x\prod_i\sum_{x_i}\eta_i(x_i)\pi(x,x_i)}
{\sum_x\prod_i\sum_{x_i}\eo(x_i)\pi(x,x_i)}\right] \, .
\end{eqnarray}
The complexity is interesting for the reconstruction problem because of the
 following 
remark:
\begin{propo}\label{propo:Stat}
Let $\Q^*$ be a distribution over $\M$ which satisfies the fixed point equation
(\ref{eq:1RSB_fixed}). Then $\Q^*$ is a stationary point of the complexity 
$\Sigma(\, \cdot\,)$. More precisely, given any consistent distribution 
 $\Q$ over $\M$,
define $\Sigma^*(t)\equiv\Sigma((1-t)\Q^*+t\Q)$. Then 
\begin{eqnarray}
\left.\frac{\de\Sigma^*}{\de t}\right|_{t=0} = 0\, .
\end{eqnarray}
\end{propo}
\prooft
This proposition is a direct consequence of Lemma \ref{remark1} in Appendix \ref{app:Variational}   
(the proof consists in explicitly computing the derivative 
of $\Sigma^*(t)$ and checking that it vanishes under the fixed point 
conditions). 
\endproof

The complexity $\Sigma$ can also be written in terms of the 
conditional distributions $Q_x(\eta) = q\eta(x)\Q(\eta)$.
Define $p(x_1,x_2)$ to be the marginal distribution of two 
neighboring variables on the tree: $p(x_1,x_2) = \pi(x_1,x_2)/q$.
Similarly, let $p(x_1\dots x_{k+1})= [\sum_x \pi(x,x_1)\dots\pi(x,x_{k+1})]/q$,
the distribution of $k+1$ variables with one common neighbor.
The complexity is then given, in terms of $Q_x(\eta)$, by:
\begin{eqnarray}
\Sigma(Q) &=& -\frac{k+1}{2}\sum_{x_1,x_2}p(x_1,x_2)
\int  W_{\rm e}(\eta_1,\eta_2)\; \de Q_{x_1}(\eta_1)
\,\de Q_{x_2}(\eta_2) +\nonumber\\
&& +\sum_{\{x_i\}}p(x_1\dots x_{k+1})\int  
W_{\rm v}(\eta_1,\dots,\eta_{k+1})\,\prod_{i=1}^{k+1} \de Q_{x_i}(\eta_i)\, ,
\end{eqnarray}
where
\begin{eqnarray}
W_{\rm e} &\equiv & - 
\log\left[\frac{\sum_{x_1,x_2}\eta_1(x_1)\eta_2(x_2)\pi(x_1,x_2)}
{\sum_{x_1,x_2}\eo(x_1)\eo(x_2)\pi(x_1,x_2)}\right] \, ,\\
W_{\rm v} &\equiv & -
\log\left[\frac{\sum_x\prod_i\sum_{x_i}\eta_i(x_i)\pi(x,x_i)}
{\sum_x\prod_i\sum_{x_i}\eo(x_i)\pi(x,x_i)}\right] \, .
\end{eqnarray}
%
%
\subsection{Implications on reconstructibility}

Experience from spin glass theory, and the  physical interpretation of the complexity,
suggests the following conjecture:
\begin{conj}\label{conj:Sigma}
Consider the reconstruction problem for the $k$-ary tree and a reversible
channel $\pi(y|x) = \pi(x|y)$.
If there exists a consistent distribution $\Q^{\rm tr}$ over $\M$, such that $\Sigma(\Q^{\rm tr})<0$,
then the reconstruction problem is solvable.
\end{conj}
Let us give here a few comments in favor of the plausibility of this
conjecture. Notice first that, if $\Q$ is trivial, then $\Sigma(\Q)=0$. Let
$\P(\M)$ denote the space of consistent probability distributions over $\M$.
Suppose that there exists $\Q^{\rm tr}$ with $\Sigma(\Q^{\rm tr})<0$.
Consider now the distribution $\Q^*\in \P(\M)$ such that the complexity is
minimal. Of course $\Sigma(\Q^*)\le \Sigma(\Q^{\rm tr})<0$ and therefore
$\Q^*$ is non-trivial. If $(i)$ $\Q^*$ is a stationary point of the complexity,
and $(ii)$ the stationary points of $\Sigma(\Q)$ in $\P(\M)$ coincide with the
solutions of the fixed point equation (\ref{eq:1RSB_fixed}), then the
existence of $\Q^{\rm tr}$ implies that the
reconstruction problem is solvable. Point $(i)$ amounts to banishing the
possibility that $\Q^*$ is on the `border' of $\P(\M)$. Point $(ii)$ is a
stronger version of Proposition \ref{propo:Stat}. 

Notice that a priori one
could formulate a similar conjecture with a $\Q^{\rm tr}$ having 
$\Sigma(\Q^{\rm tr})>0$, replacing `minimum' with `maximum' and 
`negative' with `positive' in the above. It is easy to find 
counterexamples showing that this `reverse' conjecture is false.  
The reason is  probably that the distribution $\Q^*$ maximizing 
$\Sigma(\Q)$ is on the border of $\P(\M)$, and therefore $(i)$ does not hold.

Assuming Conjecture \ref{conj:Sigma} to hold, it implies a simple variational technique
for proving that reconstruction is possible. Just consider an explicit finite-dimensional family of
distributions $\Q_{\mu}$ depending on some parameters $\mu\in{\mathbb R}^d$, and 
minimize $\Sigma(\Q_{\mu})$ over $\mu$. If the minimum is negative,
then reconstruction is possible.
We will apply the {\em variational principle} in this form in the next 
Sections. In the rest of this Section (and in Appendix
\ref{app:Variational}) we shall prove the
principle  for a special family of kernels $\pi(\,\cdot\,|\,\cdot\,)$
including the antiferromagnetic Potts channel.

We define a kernel $\pi(y|x) = \pi(x,y)$, $x,y\in\{1,\dots,q\}$
to be `frustrated' if it can be decomposed as $\pi(x,y) = \pi_*-\pih(x,y)$
where $\pi_*\in {\mathbb R}$ is a constant and $\pih(x,y)$,
$x,y\in\{1,\dots,q\}$ is a positive-definite matrix.
The antiferromagnetic Potts kernel is a particular instance of this family, with
 $\pi_* = \eps/(q-1)$, and $\pih(x,y)= |\lambda_2|\, 
\delta_{x,y}$ where $\lambda_2 = 1-q\eps/(q-1)$.

Our basic result is the following.
\begin{lemma}\label{lemma:Variational}
Let $\pi(\, \cdot\, ,\,\cdot\,)$ be a frustrated kernel and $\Q^*$
a consistent distribution over $\M$ which is not a solution of the 
associated fixed point equation (\ref{eq:1RSB_fixed}).
Then
there exists a consistent distribution $\Q$ over $\M$ such that:
\begin{eqnarray}
\left.\frac{\de\phantom{t}}{\de t}\Sigma((1-t)\Q^*+t\Q)\right|_0<0\, .
\end{eqnarray}
\end{lemma}
The proof of this statement is postponed to Appendix \ref{app:Variational}.
Here we limit ourselves to proving that it implies the desired 
principle.
\begin{propo}\label{propo:VariationalAnti}
Conjecture \ref{conj:Sigma} holds true in the case of frustrated kernels:
Let $\pi(\, \cdot\, ,\,\cdot\,)$ be a frustrated kernel, and
$\Sigma(\,\cdot\,)$ the associated complexity function. If
 there exists a consistent distribution $\Q$ over $\M$ such that 
$\Sigma(\Q^{\rm tr})<0$, 
then the reconstruction problem is solvable.
\end{propo}
\prooft
Let $\Sigma_{\min}\equiv\inf \Sigma(\Q)$, the $\inf$ being taken in $\P(\M)$. Since this 
space is subsequentially compact with respect to the weak 
topology~\cite{Shiryaev},
and $\Sigma$ is continuous with respect to this topology, there exists
a consistent distribution $\Q^*$, such that $\Sigma(\Q^*) =\Sigma_{\rm min}$.
Because of Lemma \ref{lemma:Variational}, $\Q_*$ is a solution
of Eq.~(\ref{eq:1RSB_fixed}).
Furthermore $\Sigma(\Q_*)\le \Sigma(\Q^{\rm tr})<0$ and therefore 
$\Q_*$ is non-trivial. The result is a consequence of Proposition \ref{propo:Reco}.
\endproof
%
%
\subsection{An application to Potts channels}
\label{eq:VariationalBound}
 Here we describe 
a simple family of distributions which can be used variationally
when studying the Potts channels. We will show in the next sections that, 
in spite of its simplicity, it leads to rather accurate results.

The family is indexed by a single real parameter 
$\mu\in[0,1]$. We shall denote by $\Q_{\mu}$ the corresponding distribution
and will write, with some abuse of notation, 
$\Sigma(\mu)\equiv\Sigma(\Q_{\mu})$.
The distribution $\Q_{\mu}$ attributes equal weight $1/q$ to the 
$q$ points in $\M$ denoted by $\gamma^{(x)}$, $x\in\{1,\dots,q\}$, defined as follows
\begin{eqnarray}
\gamma^{(x)}(y) = \left\{ \begin{array}{ll}
1-\mu & \;\;\;\mbox{if $y=x$},\\
\mu/(q-1) & \;\;\;\mbox{otherwise}. 
\end{array}
\right.
\label{eq:Qmudef}
\end{eqnarray}
Some calculus shows that $\Sigma(\mu) = -\frac{k+1}{2}\, w_{\rm e}(\mu)+
w_{\rm v}(\mu)$, where 
\begin{eqnarray}
w_{\rm e}(\mu) & = & -\frac{1}{q}\, A\log A-
\frac{q-1}{q}\, B\log B\, ,\\
A & = & q\left\{\frac{\eps}{q-1}+\left(1-\frac{q\eps}{q-1}\right)
\left[(1-\mu)^2+\frac{\mu^2}{q-1}\right]\right\}\, ,\\ 
B & = & q\left\{\frac{\eps}{q-1}+\left(1-\frac{q\eps}{q-1}\right)
\left[\frac{2\mu(1-\mu)}{q-1}+\frac{(q-2)\mu^2}{(q-1)^2}\right]
\right\}\, ,
\end{eqnarray}
and 
\begin{eqnarray}
w_{\rm v}(\mu) & = & -\frac{1}{q^{k+1}}\sum_{n_1,\dots, n_q}
\binom{k+1}{n_1,\dots,n_q}\, z[n]\log z[n]\, ,\\
z[n] & = & q^k\left[\frac{\eps+\mu}{q-1}-\frac{q\eps\mu}{(q-1)^2}\right]^{k+1}
\sum_{x=1}^q\left[\frac{\eps + (q-1-q\eps)(1-\mu)}{\eps + (q-1-q\eps)\mu/(q-1)
}\right]^{n_x}\, ,
\end{eqnarray} 
the first sum  being restricted to $n_1,\dots,n_q\ge 0 $ and 
$n_1+\cdot+n_q = k+1$.

Let us briefly discuss how these formulae are used in the following.
To be definite, we refer here to the ferromagnetic case, the antiferromagnetic
one being completely analogous.
Given $k$, $q$ and $\epsilon$, we compute $\Sigma(\mu)$,
and minimize it numerically for $\mu\in[0,1]$.
The largest value (more precisely, the supremum) 
of $\eps$ such that the minimum value is negative, is denoted by 
$\eps_{\rm var}(k,q)$. According to conjecture \ref{conj:Sigma}, we expect $\eps_{\rm r}(k,q)\ge\eps_{\rm var}(k,q)$.
Although we have proved it only for frustrated kernels 
(which do not include the ferromagnetic Potts channel), we shall loosely
use the term `variational bound' also in the other cases..

\begin{figure}
\begin{tabular}{cc}
\includegraphics[width=0.4\linewidth,angle=-0]{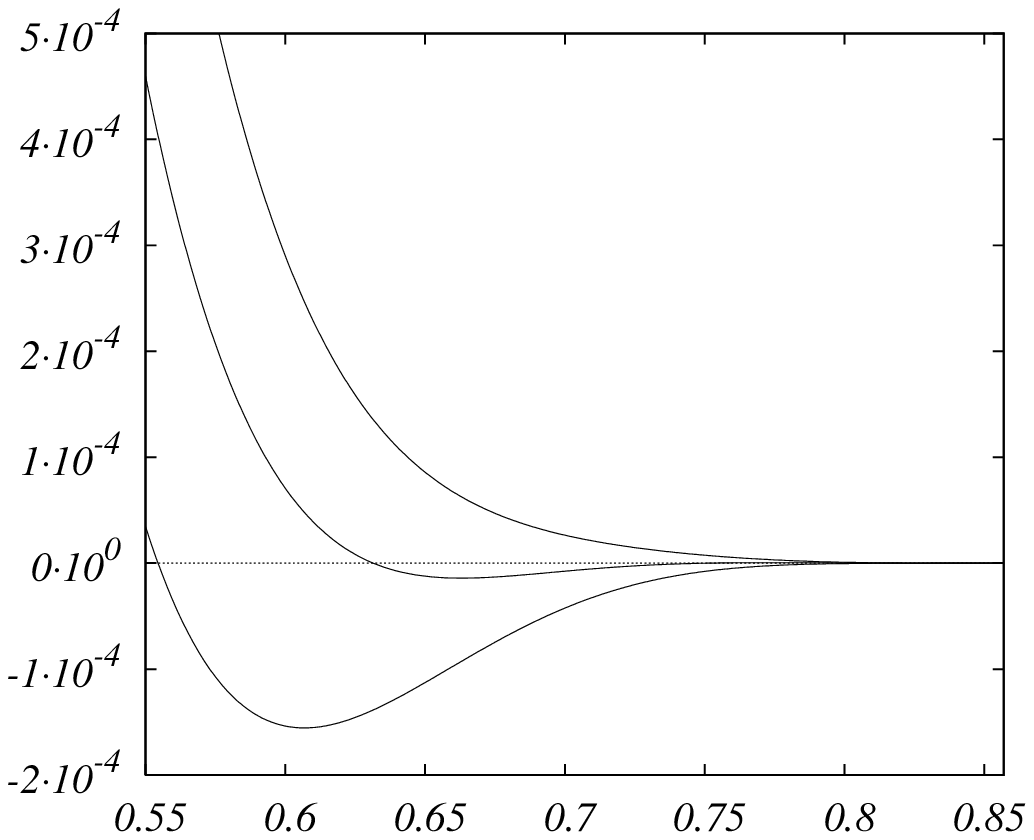}\hspace{1 cm}
\includegraphics[width=0.4\linewidth,angle=-0]{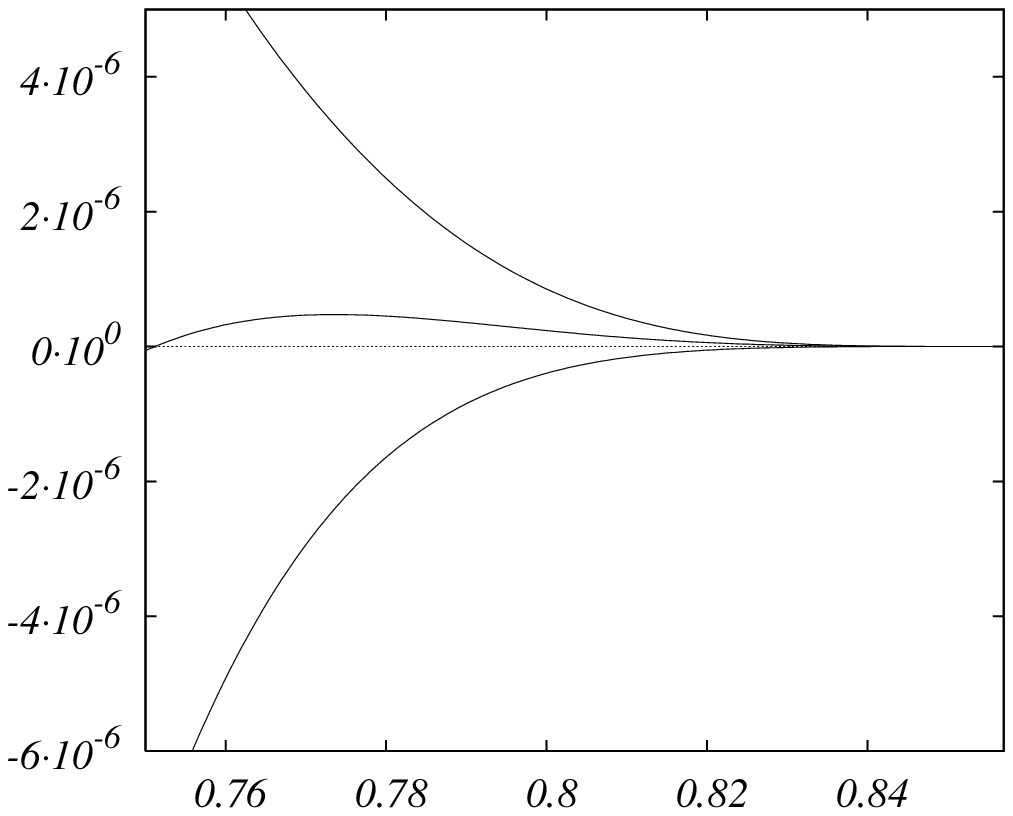}
\put(-90,-10){$\mu$}
\put(-310,-10){$\mu$}
\put(-200,70){$\Sigma(\mu)$}
\put(-420,70){$\Sigma(\mu)$}
\end{tabular}
\caption{The complexity for the ferromagnetic Potts channel with 
$k=2$ and $q=7$ within the variational ansatz $\Q_{\mu}$
described in Sec.~\ref{eq:VariationalBound}. A negative complexity
implies that the reconstruction problem is solvable. The three curves
correspond (from bottom to top) to $\eps=0.250$, $0.253$, $0.256$. 
The right plot is a zoom near  $\mu=6/7$. 
The KS threshold for $k=2,q=7$ is $\eps_{\rm KS}\approx 0.2510$. For
$\eps=0.250<\eps_{\rm KS}$,  $\Sigma(\mu)$ is negative in
the neighborhood of $\mu=6/7$. For $\eps=0.253>\eps_{\rm KS}$, 
as $\mu$ decreases from its maximum value $6/7$, $\Sigma(\mu)$ is 
first positive, but then becomes negative with a minimum for 
$\mu\approx 0.6$, implying $\eps_{\rm r}>0.253$. 
This behavior is typical of a first order phase
transition. For $\eps=0.256$, $\Sigma(\mu)$ is always positive, 
and one cannot draw any conclusion. }
\label{fig:q7}
\end{figure}

One can show that the variational bound is always at least as good 
as the KS one:
$\eps_{\rm var}(k,q)\ge \eps_{\rm KS}(k,q)$ by looking
at the behavior of  $\Sigma(\mu)$ near to $\mu = (1-1/q)$.
  By Taylor expanding $\Sigma(\mu)$
for $\mu = (1-1/q)+\delta\mu$, we obtain 
$\Sigma(\mu) = c_{k,q}(\eps) \delta\mu^4+O(\delta\mu^5)$.
Furthermore $c_{k,q}(\eps)<0$ for $\eps<\eps_{\rm KS}(k,q)$ and
 $c_{k,q}(\eps)>0$ for $\eps>\eps_{\rm KS}(k,q)$. 
In Fig.~\ref{fig:q7} we plot $\Sigma(\mu)$ for the ferromagnetic Potts channel with $k=2,q=7$,
showing that the variational  bound $\eps_{\rm var}(k,q)$ is strictly larger than the KS one.
We shall discuss in the next section for which values of $k,q$ this happens.
If the variational principle were proved for the ferromagnetic
channel, this
would prove $\eps_{\rm r}(k,q)>\eps_{\rm KS}(k,q)$ 
in these cases.

Let us notice that we do not expect the 
variational lower bound to be tight. More precisely, even minimizing it over
the space of distributions over $\M$, $\min \Sigma(\Q)$ becomes
negative only below a threshold $\eps_{\rm c}(k,q)$
with $\eps_{\rm KS}(k,q)<\eps_{\rm c}(k,q)<\eps_{\rm r}(k,q)$
(in the case where $\eps_{\rm KS}(k,q)<\eps_{\rm r}(k,q)$). Our numerical 
simulations confirm this expectation which is motivated by the physical 
interpretation of the complexity.

%
%
\section{Thresholds for the ferromagnetic Potts channel}
\label{sec:Ferro}

\begin{figure}
\includegraphics[width=0.45\linewidth,angle=0]{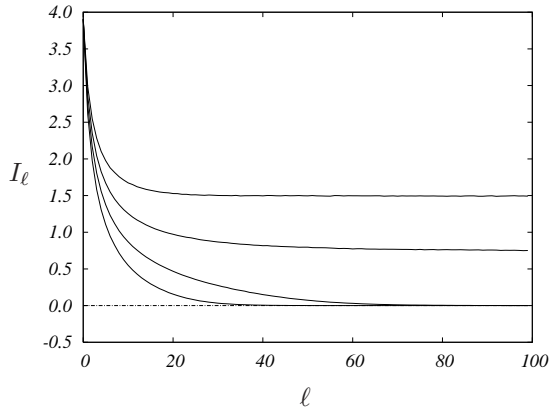}
\put(-100,-10){$\ell$}
\put(-210,75){$I_\ell$}
\caption{The information (in bits) that can be transmitted reliably through
a $k$-ary tree network of $q$-ary symmetric channels (ferromagnetic
Potts channels), as determined with the population dynamics algorithm.
Here $k=2$, $q=15$ and the noise parameter is (from top to bottom)
$\eps = 0.333082$, $0.308057$, $0.282444$, $0.256422$. 
We used populations of size $M=10^5$, and  averaged over $10$ runs.}
\label{Onerun}
\end{figure}
In order to determine reconstruction thresholds numerically,
we simulate the recursion (\ref{eq:Iteration}), 
by representing the distributions $Q^{(\ell)}_x$ through 
a large enough sample. We will estimate reconstruction to be 
possible if the sample does not concentrate, for $\ell$ large 
around the point $\eo$. 

This procedure is very similar to the  `population dynamics' method  
used to solve similar equations in spin glass theory~
\cite{Abou-Chacra,MP_Bethe}.
We work with $q$ samples (`populations')
$P^{(\ell)}_1,\dots P^{(\ell)}_q$, each containing $M$ points 
$\eta_i\in \M$, $i\in\{1,\dots,M\}$ 
(i.e. $M$ vectors $\eta_i(x)$, $x \in\{1,\dots,q\}$ with $\eta_i(x)\ge 0$ 
and $\sum_x\eta_i(x) = 1$). 
The population $P^{(\ell)}_x$ represents an i.i.d. sample 
from the distribution $Q^{(\ell)}_x$. The population
$P^{(\ell+1)}_x$, $x\in\{1,\dots,q\}$ is computed, for each 
$\ell\ge 0$ as follows.
\begin{itemize}
\item Choose $k$ iid colors $x_1,\dots,x_k$ with distribution 
$\pi(\, \cdot\,\vert x)$.
\item Choose $k$ vectors $\eta_1,\dots\eta_k$, with $\eta_i$ uniformly 
random in $P_{x_i}$.
\item Compute $\eta= \F(\eta_1,\dots,\eta_k)$ according to (\ref{eq:Site}).
\item Store this new $\eta$ in the population $P^{(\ell+1)}_{x}$, 
and repeat until the  population contains $M$ elements.
\end{itemize}
This whole cycle is repeated until the populations $P_x^{(\ell)}$ become 
stationary (by this we mean that their moments no longer depend on $\ell$)
within some prescribed accuracy.

Reconstructibility can be monitored by computing the parameters 
$I_{\ell}$ and $\Psi_{\ell}$ on in the populations 
$P_x^{(\ell)}$. If $I_\ell,\Psi_\ell\to 0$ as $\ell\to\infty$,
then we estimate that reconstruction is not possible. If they 
instead converge to a finite value, we take this value
as an estimate of $I_{\infty}$, $\Psi_{\infty}$. Figure 
\ref{Onerun} shows an example of such a calculation. 
Reconstructibility thresholds are determined by repeating the same 
experiment for several values of the channel noise $\eps$.

\begin{figure}
\includegraphics[width=0.5\linewidth,angle=0]{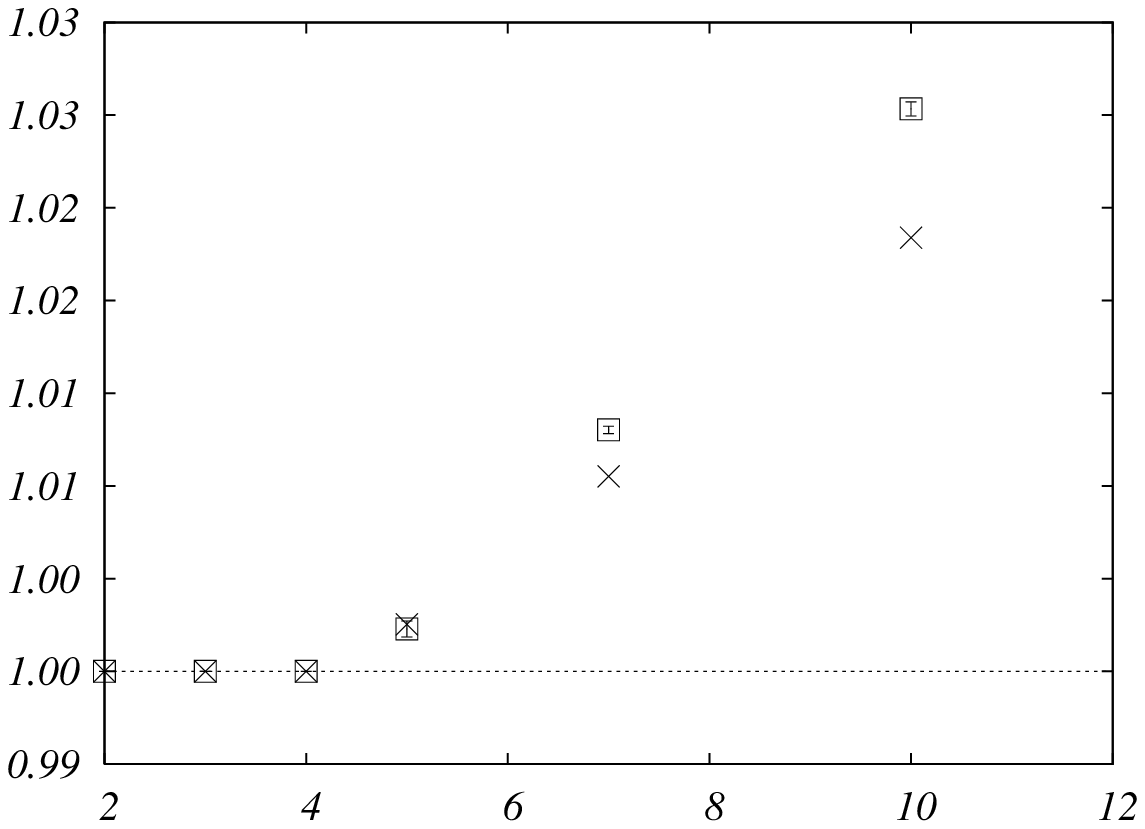}
\put(-205,90){\includegraphics[width=0.2\linewidth,angle=0]{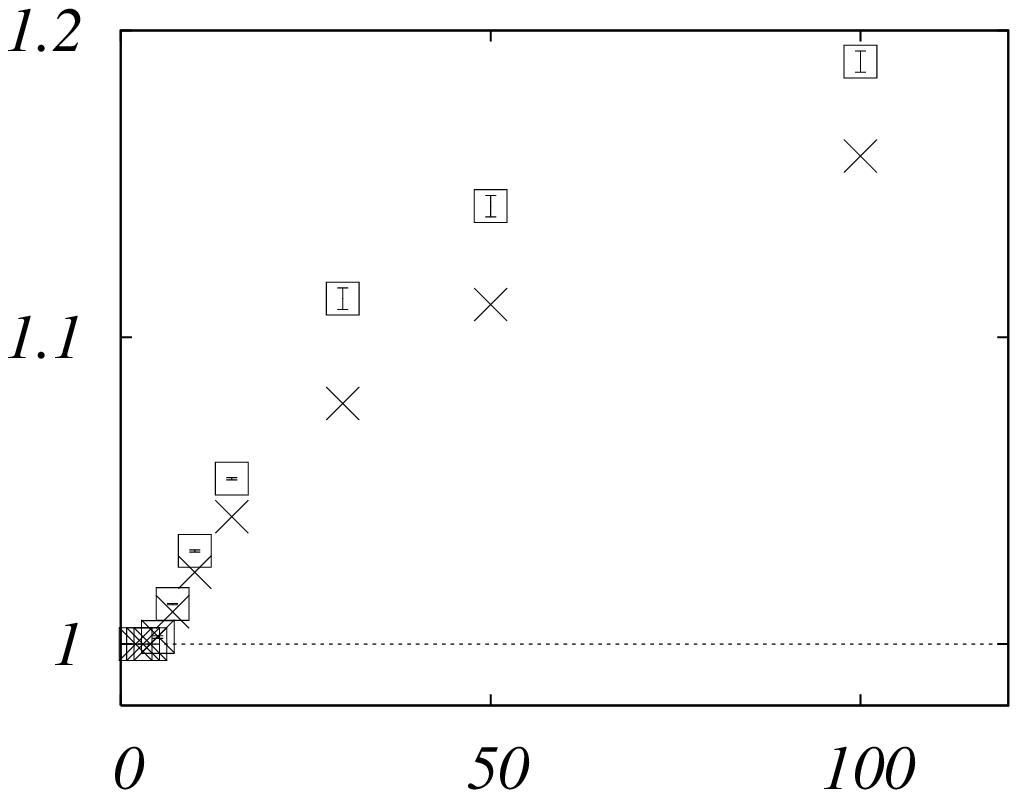}}
\put(-100,-5){$q$}
\put(-260,80){$\eps_{\rm r}/\eps_{\rm KS}$}
\caption{Ratio between the reconstructibility and the KS thresholds 
$\eps_{\rm r}(k,q)/\eps_{\rm KS}(k,q)$,
for the ferromagnetic Potts channel and $k=2$. 
Squares correspond to the numerical determination of $\eps_{\rm r}(k,q)$
and crosses to the variational lower bound $\eps_{\rm var}(k,q)$.
The inset refer to larger number of colors (up to $100$).}
\label{ExampleFig}
\end{figure}

\begin{table}
\begin{tabular}{|c|c|c|c|c|c|c|c|c|}
\hline
$q$ & $k$ & $\eps_{\rm r}$ & $\eps_{\rm KS}$ & $\eps_{\rm var}$ 
& $\eps_{\rm alg}$ &  $\eps_{\rm MP}$ &
$I_{*}$  & $\Psi_{*}$ \\
\hline
\hline
$5$ & $2$ & $0.2348(1)$ & $0.2343146$ & $0.23491$ & $---$ & 
$0.30264$ &  $0.052(5)$ & $0.0152(16)$ \\
$5$ & $3$ & $0.33881(5)$ & $0.3381198$ & $0.33887$ & $0.19047$ & 
$0.41712$ & $0.06(2)$ & $0.016(4)$  \\
$5$ & $4$ & $0.4008(1)$ & $0.4$ & $0.40081$ & $0.29046$ &  
$0.48$ & $0.06(1)$ & $0.020(4)$  \\
$5$ & $7$ & $0.4986(1)$ & $0.4976284$ & $0.49847$ & $0.41114$ &  
$0.57143$ & $0.07(1)$ & $0.020(4)$ \\
$5$ & $15$ & $0.5955(1)$ & $0.5934409$ & $0.59422$ & $0.53965$ &  
$0.65238$ & $0.14(1)$ & $0.040(8)$  \\
\hline
$7$ & $2$ & $0.25432(5)$ & $0.2510513$ & $0.25369$ & $---$ &  
$0.34577$ & $0.14(1)$ & $0.028(4)$  \\
$7$ & $4$ & $0.43325(5)$ & $0.4285714$ & $0.43250$ & $0.30769$ &  
$0.53909$ & $0.195(5)$ & $0.045(2)$\\
\hline
$10$ & $2$ & $0.2716(2)$ & $0.2636039$ & $0.26977$ & $---$ &  
$0.38325$ & $0.23(2)$ & $0.040(5)$   \\
\hline
$15$ & $2$ & $0.2881(1)$ & $0.2733670$ & $0.28472$ & $---$ &  
$0.41652$ & $0.37(3)$ & $0.053(4)$  \\
\hline
\end{tabular}
\caption{Thresholds (numerical results and bounds) for the ferromagnetic
Potts channel. 
The reconstruction threshold $\eps_{\rm r} $, 
whose numerical estimate is shown in the first column,
satisfies the rigorous bounds
$\eps_{\rm r}\ge \eps_{\rm KS}$, $\eps_{\rm r}\ge\eps_{\rm alg} $, and 
$\eps_{\rm r}\le \eps_{\rm MP}^-$.
The `algorithmic bound' $\epsilon_{\rm alg}$ is computed by analyzing
reconstruction through recursive majority along the lines of  
Ref.~\cite{MosselRec}.
The variational principle (that is not proven for this ferromagnetic channel
would imply $\eps_{\rm r}\ge \eps_{\rm var}$.
The symbol $--$ means that the corresponding bound does not provide any 
information.
}
\label{TableFerro}
\end{table}

Numerical simulations clearly show that the reconstructibility
and Kesten Stigum threshold coincide for $q=3$ and $q=4$. 
We checked this to be the case for $q=3$ and $k=2$--$7,10,15,20,30,50$,
and $q=4$ and $k=2,3,5,10,15,30$ and expect it to be the case generically, at
least for $k$ not too large. When this is the case, the order
parameters $I_{\infty}$, $\Psi_{\infty}$ decrease continuously and vanish at 
$\eps_{\rm r}(k,q)=\eps_{\rm KS}(k,q)$.

For $q\ge 5$ we always find $\eps_{\rm r}(k,q)>\eps_{\rm KS}(k,q)$.
In these cases $I_{\infty}(\eps)\downarrow I_*>0$,  
$\Psi_{\infty}(\eps)\downarrow \Psi_*>0$ as $\eps\uparrow \eps_{\rm r}(k,q)$.
In spin glass language, the transition is discontinuous:
we refer to next Section for some illustrations. 
We report our numerical results in Table 
\ref{TableFerro}. This table also contains
 the  variational lower bound $\eps_{\rm var}(k,q)\le\eps_{\rm r}(k,q)$, 
as well as the upper bound derived in
 \cite{MosselPeres01}:
$\eps_{\rm r}(k,q)\le \eps_{\rm MP}^{+}(k,q)$,
where
\begin{eqnarray}
\eps_{\rm MP}^{\pm}(k,q) = (q-1)\frac{(2-q+2kq)\mp\sqrt{(2-q+2kq)^2
-4k(k-1)q^2}}{2kq^2}\, .
\end{eqnarray}

In Fig.~\ref{ExampleFig} we plot the thresholds as a function of
$q$ for $k=2$.
%
%
\section{Thresholds for the antiferromagnetic Potts channel} 
\label{sec:AntiFerro} 

\begin{table}
\begin{tabular}{|c|c|c|c|c|c|c|c|c|c|}
\hline
$q$ & $k$ & $\eps_{\rm r}$ & $\eps_{\rm KS}$ & $\eps_{\rm var}$ 
& $\eps_{\rm alg}$ & $\eps_{\rm MP}^-$ 
& $I_{*}$  & $\Psi_{*}$ & $\Sigma_{*}$ \\
\hline
\hline
$4$ & $8$ & $0.99953(4)$ & $--$ & $--$ & $--$ & $0.91552$ & 
$1.56(4)$ & $0.56(1)$ & $0.026(3)$\\
$4$ & $9$ & $0.9908(4)$ & $1$ & $0.99298$ & $--$ & $0.90717$ & 
$1.31(2)$ & $0.47(2)$ & $0.009(1)$\\ 
$4$ & $10$ & $0.9820(8)$ & $0.9871708$ & $0.98304$ & $--$ & $0.9$ &
$1.2(2)$ & $0.42(4)$ & $0.005(4)$\\
$4$ & $11$ & $0.9725(3)$ & $0.9761335$ & $0.97363$ & $0.99736$ & $0.89376$ & 
$1.07(5)$ & $0.39(1)$ & $\lesssim 0.005$\\
$4$ & $12$ & $0.9643(3)$ & $0.9665063$ & $0.96498$ & $0.98946$ & $0.88826$ &
$0.26(3)$ & $4.2(5)$ & $\lesssim 0.005$ \\ 
$4$ & $15$ & $0.9431(3)$ & $0.9436492$ & $0.94338$ & $0.96903$ & $0.875$ &
$0.5(1)$ & $0.16(3)$ & $\lesssim 0.001$ \\
$4$ & $18$ & $0.9267(2)$ & $0.9267766$ & $0.92686$ & $0.95264$ & $0.86502$ &
$0.3(1)$ & $0.11(4)$ & $\lesssim 0.001$ \\
\hline
$5$ & $13$ & $0.99741(5)$ & $--$ & $0.99982$ & $--$ & $0.92308$ & 
$1.76(4)$ & $0.59(1)$ & $0.042(5)$\\
$5$ & $14$ & $0.9932(1)$ & $--$ & $0.99555$ & $--$ & $0.91916$ &
$1.7(1)$ & $0.54(2)$ & $0.03(1)$\\
$5$ & $15$ & $0.9888(1)$ & $--$ & $0.99092$ & $--$ & $0.91561$ &
$1.48(5)$ & $0.48(2)$ & $0.03(1)$\\
$5$ & $20$ & $0.9685(3)$ & $0.9788854$ & $0.96991$ & $0.98581$ & $0.90177$ &
$1.1(5)$ & $0.36(2)$ & $0.01(1)$\\
\hline
$6$ & $17$ & $0.999924(5)$ & $--$ & $--$ & $--$ & $0.93482$ & 
$2.20(4)$ & $0.667(15)$ & $0.095(5)$  \\ 
$6$ & $20$ & $0.9932(3)$ & $--$ & $0.99546$ & $--$ & $0.92792$ &
$1.87(6)$ & $0.569(15)$ & $0.04(2)$\\ 
\hline
\end{tabular}
\caption{Thresholds (numerical results and bounds) for the antiferromagnetic
Potts channel.
The reconstruction threshold $\eps_{\rm r} $, 
whose numerical estimate is shown in the first column,
 satisfies the rigorous bounds
 $\eps_{\rm r}\le \eps_{\rm KS}$ (from \cite{KestenStigum2}), 
$\eps_{\rm r}\le\eps_{\rm alg}$ (cf. \cite{MosselRec}), 
$\eps_{\rm r}\le\eps_{\rm var} $ (from Proposition \ref{propo:VariationalAnti}), 
and 
 $\eps_{\rm r}\ge \eps_{\rm MP}^-$ (from \cite{MosselPeres01}).
 The symbol $--$ means that the corresponding bound does not 
provide any information.}
\label{TableAnti}
\end{table}
\begin{figure}
\includegraphics[width=0.45\linewidth,angle=0]{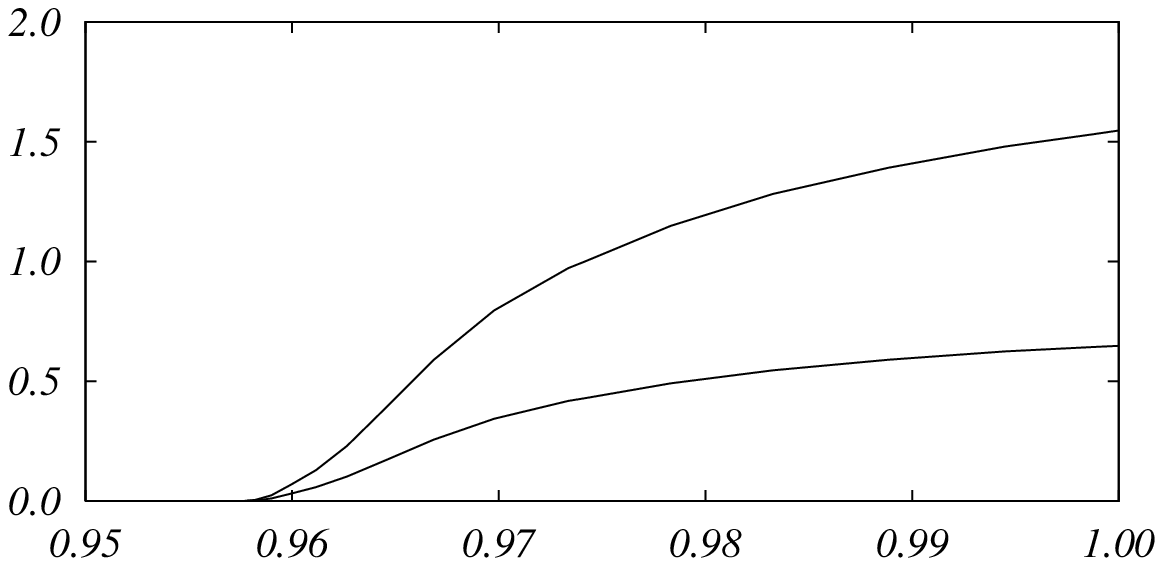}
\hspace{1cm}
\includegraphics[width=0.45\linewidth,angle=0]{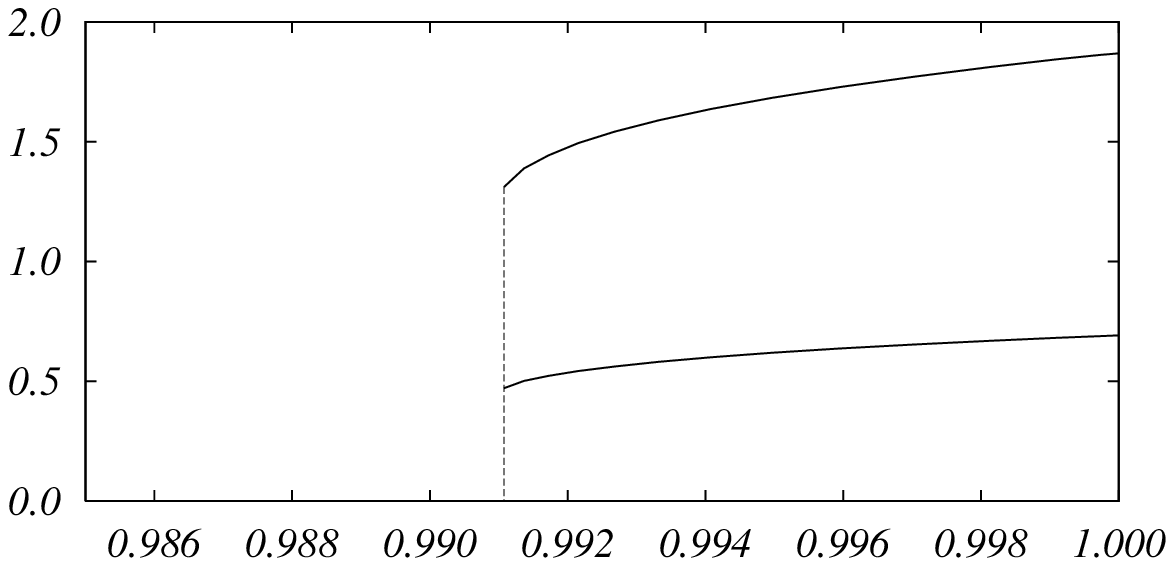}
\put(-465,95){\includegraphics[width=0.475\linewidth,angle=0]{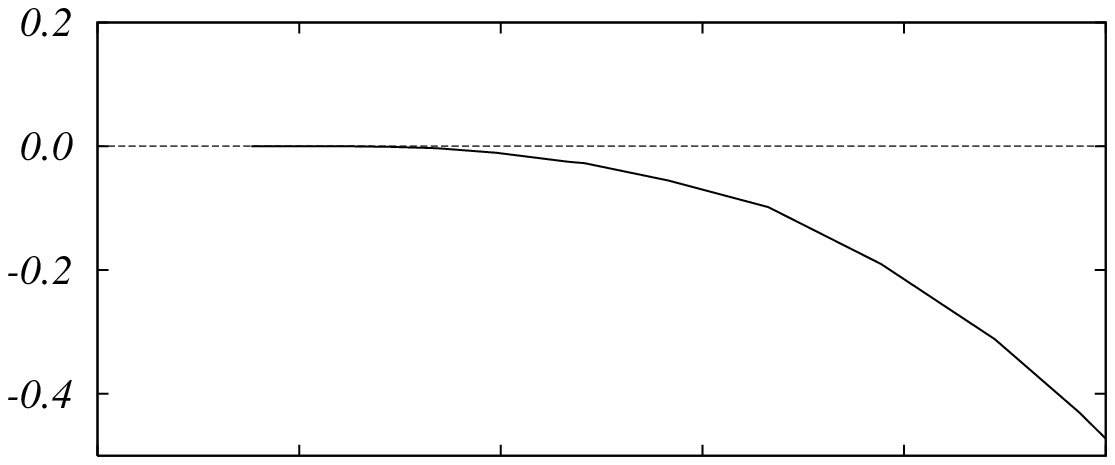}}
\put(-221,95){\includegraphics[width=0.475\linewidth,angle=0]{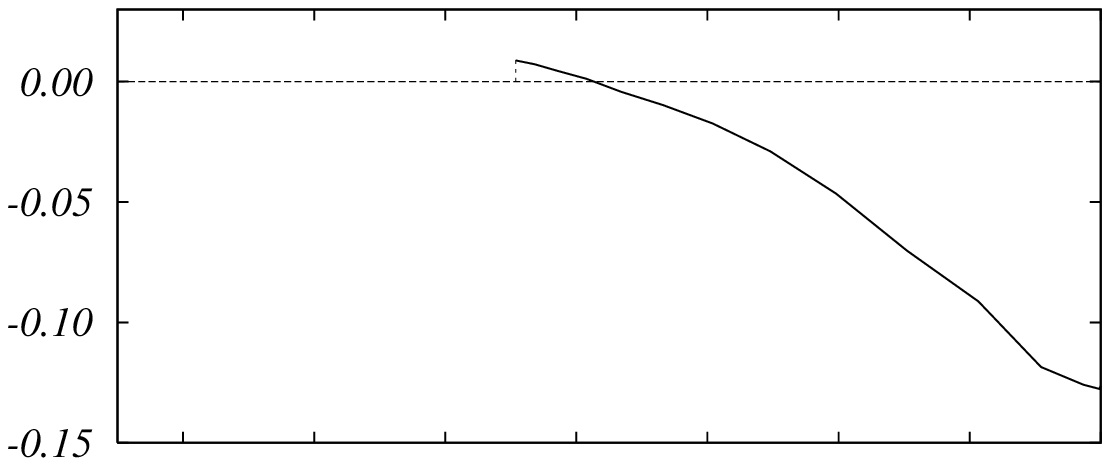}}
\put(-100,-10){$\eps$}
\put(-350,-10){$\eps$}
\put(-234,50){$\Psi_{\infty},I_{\infty}$}
\put(-480,50){$\Psi_{\infty},I_{\infty}$}
\put(-225,138){$\Sigma_{\infty}$}
\put(-474,138){$\Sigma_{\infty}$}
\caption{Asymptotic complexity $\Sigma_{\infty}$, information capacity 
$I_{\infty}$ and conditional variance $\Psi_{\infty}$ as a function
of the noise parameter for the antiferromagnetic Potts channel.
On the left: typical continuous reconstructibility transition,
$q=3$, $k=6$. On the right: typical discontinuous transition,
$q=4$, $k=9$. }
\label{DisContinuousFig}
\end{figure}

In Table \ref{TableAnti} we present our numerical results for the 
reconstruction thresholds of the antiferromagnetic 
Potts channel in the cases in which it differs from $\eps_{\rm KS}$, together with the
bounds. 

One distinctive feature of this channel is that, even
in the limit $\eps\to 1$ reconstruction may be impossible. For any 
given $q\ge 3$ reconstruction becomes possible only for 
$k\ge k_*(q)$. Numerically we found $k_*(3) = 5$, $k_*(4)=8$, $k_*(5)=13$,
$k_*(6)=17$.  In fact the case $\eps=1$ has a special 
interest. In this case the configuration produced by the broadcast process 
is  distributed according to the free boundary
Gibbs measure for proper colorings of the (infinite) tree $\Tree_k$.
Our numerical results imply that this measure is extremal only for 
$k<k_*(q)$, with $k_*(q)$ as above. Using the variational principle
(which in this case is proved, cf. Proposition \ref{propo:VariationalAnti}), 
we can show that $k_*(3)\le 5$,  $k_*(4)\le 9$, $k_*(5)\le 13$,
$k_*(6)\le 17$\dots

For $q=3$ we found the reconstructibility threshold to coincide
always with the KS threshold. This was checked for
$k=4$--$7$, $10$, $20$. The parameters $I_{\infty}$ and $\Psi_{\infty}$ are 
continuous functions of $\eps$ vanishing at $\eps_{\rm KS}$.
An example is provided in Fig.~\ref{DisContinuousFig}, left frame.
For $q\ge 4$ and $k\ge k_*(q)$ the transition at the reconstructibility
threshold is discontinuous, cf. Fig.~\ref{DisContinuousFig}, right frame.
Table   \ref{TableAnti} gives the values of $I_{*}$, $\Psi_{*}$
and $\Sigma_{*}\equiv \lim_{\eps\to \eps_r}\Sigma(\Q^{(\infty)})$
(in the ferromagnetic channel, this number is so small that it cannot be measured reliably in the numerics).
Most of the remarks made for the ferromagnetic channel apply to this case.

%
%
\section{Relation to spin glass theory} 
\label{se:spinglass}

In this section we explore the link between reconstruction and spin glass
theory. 
For simplicity we keep to the Potts channel, but the discussion
can be easily generalized. 
Consider a configuration $\uY^{L}$ generated by the
broadcast along a finite rooted tree $\Tree_k(L)$ with $L$ 
generations, starting from a  uniformly random symbol in 
$\{1,\dots,q\}$ at the root.
As we saw in the introduction, $\uY^L$ is an equilibrium 
configuration of the Potts model with free boundary conditions
on $\Tree_k(L)$, i.e. is distributed according to the
Boltzmann law for the energy function (\ref{eq:PottsEnergy}).
The coupling $J$ of this model is given by 
$e^{-\beta J}=\frac{\eps}{(q-1) (1-\eps)}$, 
and is ferromagnetic (resp. antiferromagnetic) if $J>0$ (resp $J<0$).

Once the  broadcast process has fixed the variables on
the boundary at distance $L$ from the root, the reconstruction
problem can be phrased in terms of the conditional distribution
$\prob\{X_0=x|\uX_{L}=\ux_{L}\}$. 
The distribution of the first $L-1$ generations given the received 
symbols, $\prob\{\uY^{L-1}|  \uX_{L}=\ux_{L}\}$,
is also given by Boltzmann law for the energy function (\ref{eq:PottsEnergy}).
However the boundary condition is now given by the received symbols.
One fundamental reason why reconstruction is related to a spin glass 
problem is that this boundary condition tends
to frustrate the system, in the sense of creating conflicting constraints.

It is well known that on trees, frustration comes only through the choice 
of boundary conditions. Here we discuss the spin glass phase induced in the 
Potts model on the tree by various possible choices. We shall first show how a
`naive' choice of boundary conditions leads to a simple replica symmetric
recursion relation. Then we show how well-chosen self-consistent boundary
conditions lead to the correct 1RSB fixed point equation of 
reconstruction.
Finally we discuss the explicit realization of the corresponding spin 
glass
model as a model of Potts spins on a random graph (not a tree).
%
%
\subsection{Boundary conditions with independent spins}
As before, we call $\uX_{L}$ the set of all spins in the $L^{\rm th}$
generation. A boundary condition (BC) is a probability distribution  on 
these spins. One first possibility is when spins on the boundary are
independent random variables: a given spin $X_i$, $i\in V_{L}$ takes 
value $x_i$ with probability $\eta_i(x_i)$. 
The overall distribution is therefore $\prod_{i \in V_{L}}\eta_i(x_i)$. 
Once a set of $\eta_i(\, \cdot\, )$'s is given, the partition function of 
the tree is obtained as:
\begin{equation}
Z_L(\{\eta_i\}_{i\in V_L})=\sum_{\uy^L} \prod_{(i,j)\in \Tree_k(L)} 
\pi(x_i,x_j) \prod_{i\in V_L} \eta_i(x_i)\, ,
\label{Zdef}
\end{equation}
and the Boltzmann distribution is 
\begin{equation}
\prob^L_{\{\eta_i\} }(\uy^{L}) =\frac{1}{Z_L(\{\eta_i\}}\; 
\prod_{(ij)\in \Tree_k(L)} 
\pi(x_i,x_j) \prod_{i\in V_L} \eta_i(x_i)\, .\label{eq:BoltzmannBC}
\end{equation}

We have still the freedom of chosing the $\eta_i(\, .\,)$.
One  simple possibility would be to take them identical:
$\eta_i(\, \cdot\, )= \overline{\eta}(\, \cdot\, )$ 
\cite{Peruggi83,Peruggi84a,Peruggi84b}, but in order to have a
 disordered and frustrated problem one can choose to sample 
the $\eta_i$'s independently  from a {\it symmetric} distribution 
$P^{(0)}(\eta)$. This definition of a spin glass problem on a tree 
was adopted for instance in \cite{Chayes86} in the Ising case ($q=2$),
where each of the boundary spins was fixed to 
$\pm 1$ independently. In our formulation, this corresponds to the choice
$P^{(0)}(\eta)=\frac{1}{2}\left(\delta\left[\eta(\, \cdot\,),\delta_{+1}\right]
+\delta\left[\eta(\,\cdot\,),\delta_{-1}\right]\right)$.
 
The recursive procedure for merging rooted trees applies in the same way 
as in Sec.~\ref{se:merging}. Consider the marginal distribution on the
first $L-\ell\le L$ generations, $\prob(\uy^{L-\ell})$. It is clear that
this has the same form as in Eq.~(\ref{eq:BoltzmannBC}),
with some new $\eta'_i(\,\cdot\, )$, $i\in V_{L-\ell}$.
When one generate BCs randomly as described above, the $\eta'_i$ are 
iid random variables with common distribution 
$P^{(\ell)}(\eta)$. A little thought  shows
that $P^{(\ell)}(\eta)$ is related to the one in the shell just 
above by:
\begin{equation}
P^{(\ell+1)} (\eta)=
\int 
\delta\left[\eta- \F(\eta_1,\dots,\eta_k)\right]\;\; \prod_{i=1}^k 
\de P^{(\ell)}(\eta_i)
\, .
\label{eq:iterRS}
\end{equation}
Notice that in each shell, $P^{(\ell)}$ is symmetric.

The marginal distribution at the root of the tree $\Tree_k(L)$, 
under the Boltzmann law (\ref{eq:BoltzmannBC}),
is a random variable with distribution $P^{(L)}(\eta)$.
In this model, the existence of a spin glass phase is characterized by
a non trivial limit of $P^{(L)}(\eta)\to P^{(\infty)}(\eta)$ as $L\to\infty$.
Such a limit solves  a  fixed point equation corresponding to
(\ref{eq:iterRS}). 

The reader will notice that Eq.~(\ref{eq:iterRS}) is similar to 
the reconstruction equation (\ref{eq:1RSB}), with one crucial difference:  
the `reweighting' factor $z(\{\eta_i\})$ in
(\ref{eq:1RSB}) is absent here. In the spin-glass jargon, 
Eq.~(\ref{eq:iterRS}) is the  `replica symmetric' (RS) equation,
while Eq.~(\ref{eq:1RSB}) is the 1RSB
equation with Parisi parameter $m=1$.

We want to argue that  the model defined by Eq.~(\ref{eq:BoltzmannBC}), with iid $\eta_i$'s
in not a `good' model of spin glass on a tree.
Technically this is seen from the fact that its glass
phase is  a RS one, while the spin glass models on graphs with loops typically show RSB.
Fundamentally, the drawback of this model is precisely
that it neglects correlations between spins on the boundary. 
As we will discuss in Sec.~\ref{Bethe_lattice}, such correlations are necessary in order
to study the existence of many pure states, a distinguished mark of spin glasses on graphs
with loops.

A minimalistic way of introducing such correlations is
to keep uniquely those correlations induced by the tree itself;
this is precisely what is done in reconstruction,
as we now discuss.
%
%
\subsection{Self-consistent boundary conditions}
It is clear that the broadcast/reconstruction process generates correlated BCs.
By this we mean that the conditional distribution
$\prob(\uY^{L-1}=\uy^{L-1}|\uX_L)$ has still the form 
(\ref{eq:BoltzmannBC})
but the $\eta_i$ are no longer independent. 
More explicitely, the $\eta_i$'s, $i\in  V_L$ 
have distribution:
\begin{equation}
\prob(\{\eta_i\}_{i\in V_L})= \frac{1}{\Xi_L}\, 
Z_L(\{\eta_i\}_{i\in V_L}) \,
\prod_{i\in V_L} \QQ^{(0)}(\eta_i) \ ,
\label{scBC}
\end{equation} 
where $Z(\{\eta_i\})$ is the partition function 
(\ref{Zdef}) of the tree  $\Tree_k(L)$ with  BC $\{\eta_i\}$,
and $\QQ^{(0)}(\eta)$ is the uniform distribution on the $q$ 
`corners' of the simplex 
$\eta(x)=\delta_{x,r}$, $r\in\{1,\dots,q\}$.

We can analyze the system with BC 
(\ref{scBC}) as we did in the previous section for uncorrelated BCs.
Consider, as before, the marginal distribution of the first $L-\ell\le L$
generations of the tree. It also has the form 
(\ref{eq:BoltzmannBC}) and
the new $\eta_i$'s at distance $\ell$ retain the same correlation structure. 
Their joint distribution is
\begin{equation}
\prob(\{\eta_i\}_{i\in V_\ell})= \frac{1}{\Xi_{\ell}}\, 
Z_\ell(\{\eta_i\}_{i\in V_\ell}) \,
\prod_{i\in V_{\ell}} \QQ^{(\ell)}(\eta_i) \, .
\label{scBC2}
\end{equation}
Finally, the distribution
$\QQ^{(\ell+1)}$ is related to $\QQ^{(\ell)}$ through the recursion 
(\ref{eq:1RSB}) that
we found when discussing reconstruction, with the correct reweighting 
factor. Since we took $\QQ^{(0)}(\eta) = \Q^{(0)}(\eta)$,
this implies $\QQ^{(\ell)}(\eta) = \Q^{(\ell)}(\eta)$ for any
$\ell\ge 0$.

It is interesting to define the spin glass problem on the tree associated 
with the non-trivial fixed point of Eq.~(\ref{eq:1RSB_fixed}). 
This just amounts to  generating
the BCs from (\ref{scBC}) using $\QQ^{(0)}=\Q^{*}$. This problem has the
virtue of being statistically translation 
invariant~\footnote{Provided one replaces the rooted tree
(with a root of degree $k$) with a regular Cayley tree 
(with all the vertices  of degree $k+1$).}
(although, for any given realization, the resulting Gibbs measure is 
not translation invariant).
In particular the properties of a spin don't depend on its distance to the 
root.
%
%
\subsection{Spin glass on the Bethe lattice}
\label{Bethe_lattice}
While the previous definition of a spin glass on a tree is perfectly
correct, it is clear that a lot of the physics has been put into the choice 
of the BC distribution. This is necessary because of the crucial role of BCs 
on trees. An alternative definition of the Bethe lattice spin glass, proposed 
in \cite{MP_Bethe}, is to use, instead of trees, random graphs which have a 
tree structure on finite length scales. Let us consider for instance the 
problem of $N$ Potts spins on the vertices of a random regular graph 
$\cG_N$ with  degree
$k+1$, with pairwise interactions given by the kernel $\pi(x,y)$
The partition function of such a model is
\begin{equation}
Z=\sum_{x_1,\dots,x_N} \prod_{(i,j)\in \cG_N} \pi(x_i,x_j)
\label{eq:randGr} 
\end{equation}
In any finite neighborhood of a randomly chosen node $i$, the local 
structure~\footnote{By this we mean the subgraph within any fixed distance 
from $i$. The property described here can also be phrased in  
terms of {\em local weak convergence}~\cite{AldousSteele}} of $\cG_N$ 
is (with high probability) the one of a regular tree with degree $k+1$. 
In fact, the shortest loop through $i$
is typically of size $\log N$ which diverges when $N\to\infty$. 
This setting is interesting for two reasons:
$(i)$ Loops, although large, can create some frustration; 
$(ii)$ The system is approximately homogeneous (unlike on a regular tree,
where vertices on the boundary have a neighborhood very 
different from the others).

Spin glasses on random lattices with a local tree-like structure have been 
the object of many studies in recent times. The cavity method of
\cite{MP_Bethe,MP_Bethe_T0} is an iterative procedure which exploits the
tree-like structure. Here we shall just mention some of its main results 
without justification, the aim being to clarify the correspondence
between the spin glass model on a random graph and the reconstruction 
problem.

\begin{figure}
\includegraphics[width=0.25\linewidth,angle=0]{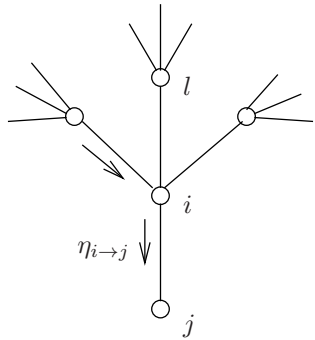}
\put(-50,-5){$j$}
\put(-50,40){$i$}
\put(-50,85){$l$}
\put(-90,25){$\eta_{i\to j}$}
\caption{The basic cavity recursion.}
\label{fig:CavityRecursion}
\end{figure}
In the cavity method one first considers the graph, rooted in a site $i$,
obtained by cutting the edge $(i,j)$ to one of its neighbors, cf. 
Fig.~\ref{fig:CavityRecursion}. The 
marginal distribution of the root, $\eta_{i \to j}(x_i)$, 
with respect to the model on the `amputated' graph, is then 
written in terms of the distributions $\eta_{l\to i}(x_l)$ 
where $l$ are the  neighbors of $i$ different from $j$. It is possible to write 
such a recursion only if the variables $x_l$, 
in the absence of the edges $(l,j)$, become  uncorrelated in the large $N$ 
limit. Such a property is made possible by the local 
tree-like structure,  but  also requires 
a fast decay of correlations in the graph.
This is expected to happen either when the system admits 
a single Gibbs state, or when the Boltzmann measure is restricted 
to a pure (extremal) Gibbs state~\footnote{The definition of 
extremal Gibbs state on a {\em finite} graph goes beyond the scope of this
paper.}.  In 
the first case, the problem is described by a unique distribution $\eta(x)$, 
and $\eta_{i \to j}=\eta$ for all directed edges $i\to j$. 
This distribution is a fixed
point of (\ref{eq:Site}), satisfying thus $\eta=\F(\eta,\dots,\eta)$. It 
is called the 'paramagnetic' or 'liquid' phase. 
In the case where there exist
several pure states, the recursion holds when the measure is restricted to 
one pure state $\alpha$: on a given (large) graph one could thus generate a 
set of `messages' $\eta_{i\to j}^\alpha(x_i)$ for each state $\alpha$. Notice 
that, for a given $\alpha$, the messages now depend explicitely on the 
edge: the measure is no longer uniform, but it is modulated. The 1RSB 
cavity method assumes that there exist exponentially 
many such pure states, the number $\cN(f)$ of states with free energy density 
$F^\alpha/N=f$ is written in terms of the complexity function $\Sigma(f)$ as 
$\cN(f)=\exp(N \Sigma(f))$. In such a case one can perform a statistics in the space of 
pure
states, by introducing, for each edge $(i,j)$, the probability $R_{i\to
j}(\eta)$ that the message $\eta_{i\to j}^\alpha=\eta$, when $\alpha$ is 
chosen
randomly with a weight proportional to the total Boltzmann weight of state
$\alpha$. After performing this average over states the various edges 
become
again equivalent, and one finds that the distribution $R_{i\to 
j}(\eta)=\Q^*$
satisfies exactly the 1RSB fixed point equation (\ref{eq:1RSB_fixed}). So
there exists a 1RSB glass phase if and only if this equation has a non 
trivial
symmetric solution. Notice that this equation can also have other
non-symmetric solutions. For instance in the case of the ferromagnetic 
Potts
channel, at low enough temperature there is a solution where $Q^*$ is 
peaked
on a $\eta$ with a ferromagnetic bias, but it does not satisfy the 
symmetry
property that we impose for the study of the glass state. In such a system 
the
glass solution exists, but it is not realized on a random graph: the 
system
will transit to a ferromagnetic phase. On the contrary in some other cases 
the
glass phase will be realized. For instance we expect this to be the case 
for
the antiferromagnetic Potts model on the random graph \cite{Mulet02,Brauenstein03}.

The tree  reconstruction problem on the one hand, and the spin glass on a
random graph on the other, thus naturally lead to the same equations.
Some aspects of this correspondence call for a better understanding. 
Consider the model on the random graph defined in 
Eq.~(\ref{eq:randGr}). Let us suppose that it has several pure states, and that
the 1RSB cavity solution of the problem is correct. Now isolate around an
arbitrary point the set of all its neighbors up to distance $\ell$.
Generically it is a tree $\Tree_k(\ell)$. The vertices outside this tree create
some boundary condition on the leaves of this tree, depending 
on the pure state $\alpha$ that we are considering. 
We have found  that
the statistics of these BC on the pure states corresponds to the statistics
of the boundaries in the broadcast/reconstruction, and both are described
by the distribution $\Q^*$.
 In spin glass theory (within 1RSB) one can count
the pure states through the computation of the complexity function
$\Sigma(f)$. It would be very interesting to have an
interpretation of this function in terms of the reconstruction problem.
%
%
\section{Generalizations}
\label{se:gene}

So far we have focused on $k$-ary trees whose links corresponds 
to identical copies of the same $q$-ary channel satisfying the symmetry 
condition $\pi(y|x) = \pi(x|y)$. However, none of these hypotheses is  
crucial to our approach. In this section we define a considerably more
general context, and sketch how to adapt the above formalism to this case.
Some cases of broadcast through non regular trees, or with  asymmetric channels have been considered
for instance in  \cite{EvaKenPerSch,Mossel01,MosselPeres01,Martin03}. The present formalism encompasses
all these cases and generalizes them to broadcast through hypergraphs.

\begin{figure}
\includegraphics[width=0.3\linewidth,angle=-0]{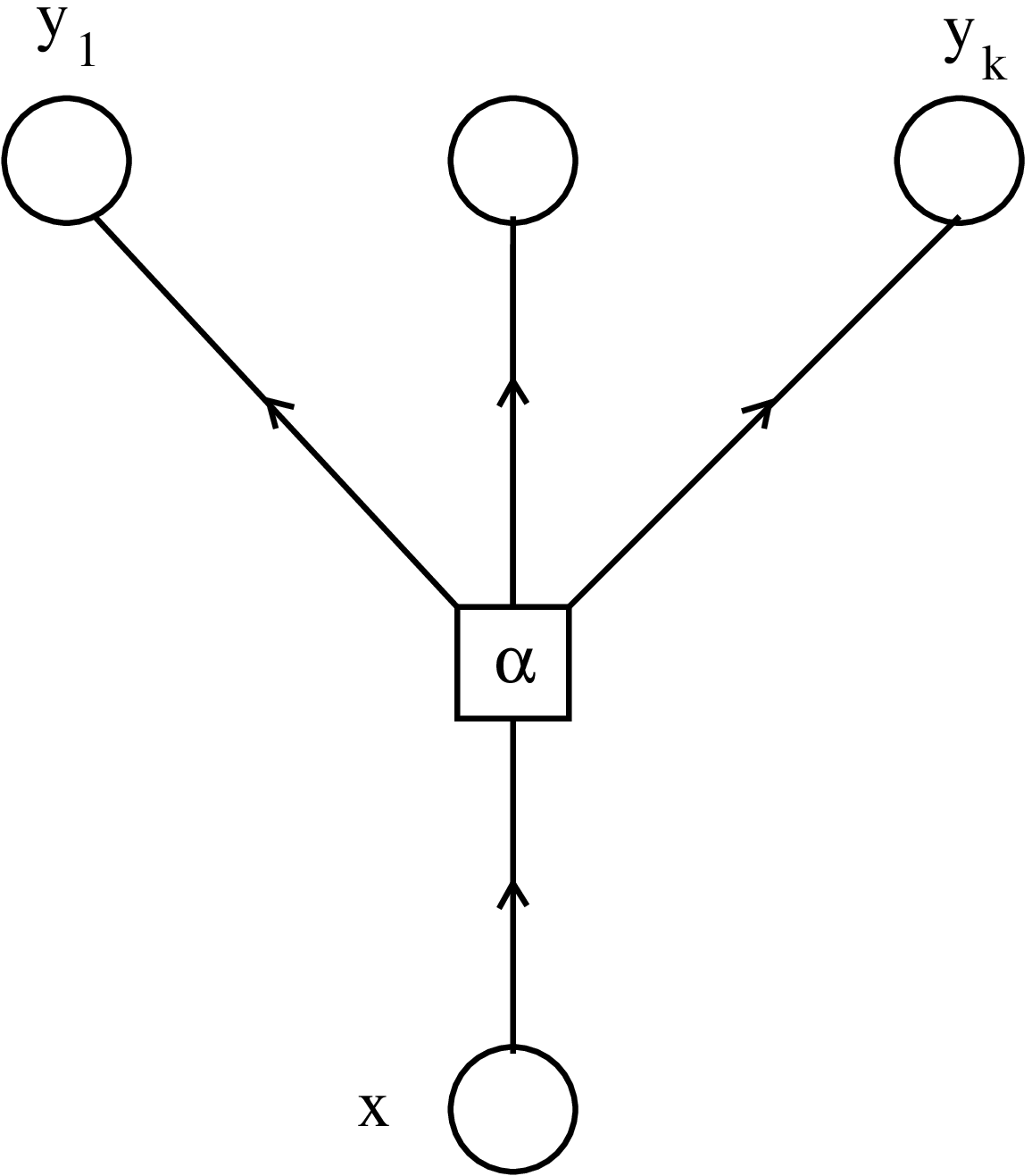}\hspace{1 cm}
\includegraphics[width=0.6\linewidth,angle=-0]{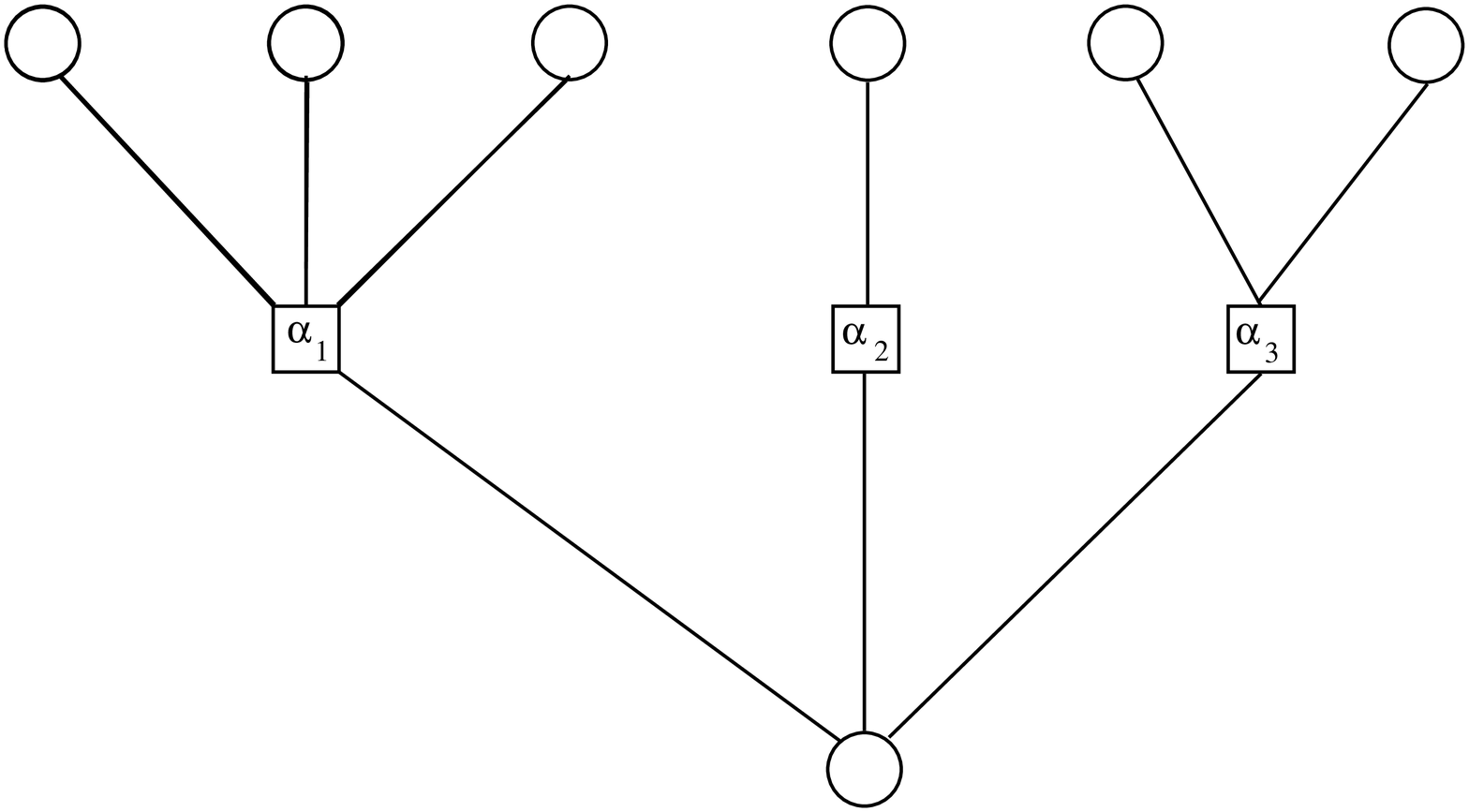}
\caption{Left: a function node representing the  `one-to-$k$' channel $\pi^{(\alpha)}(y_1,\dots,y_{k} |x)$.
Right: An example of a small random tree network, with $l_0=3$ }
\label{fig:fnode}
\end{figure}

We consider a finite set of kernels 
$\{\pi^{(\alpha)}(\, \cdot\, |\,\cdot\,)\, ;\, \alpha=1,\dots,n\}$,
each kernel describing a one-to-many communication channel.
For each $x\in\{1,\dots,q\}$, and each $k_{\alpha}$-uple
$y_1,\dots, y_{k_\alpha}$,
$\pi^{(\alpha)}(y_1,\dots,y_{k_{\alpha}} |x)$ gives is the probability
that users $1,\dots,k_\alpha$ receive outputs $y_1,\dots,y_{k_{\alpha}}$
if the channel input was $x$. These kernels must
satisfy the conditions
\begin{eqnarray}
\pi^{(\alpha)}(y_1,\dots,y_{k_{\alpha}} |x)\ge 0\, ,
\;\;\;\;\;\;\; \sum_{y_1,\dots,y_{k_{\alpha}}}
\pi^{(\alpha)}(y_1,\dots,y_{k_{\alpha}} |x) = 1\, ,
\end{eqnarray}
and $x$ is called the parent of $y_1,\dots,y_{k_{\alpha}}$. Such `one-to-$k$' communication 
channels can be represented graphically
using factor nodes of degree $k+1$, cf. Fig.~\ref{fig:fnode}. 

Next, we define a {\em random} tree network ensemble depending
on two probability distributions $q_{\alpha}$, $\alpha\in\{1,\dots,n\}$ 
($q_{\alpha}\ge 0$, $\sum q_{\alpha}=1$) and
$p_l$, $l\ge 0$ ($p_{l}\ge 0$, $\sum p_{l}=1$).
One (infinite) random network $\Tree$ 
from this ensemble is generated as follows
starting from the root $0$.
\begin{itemize}
\item Draw an integer $l_0$ with distribution $p_l$. This is the degree
of the vertex $0$
\item For each $a\in\{1,\dots, l_0\}$, draw $\alpha_a$ independently with 
distribution $q_{\alpha}$, and attach a channel of type $\alpha$
to the vertex $0$. The root will transmit through such channels.
\item For each  $a\in\{1,\dots, l_0\}$, and each $i_a\in\{1,\dots,
k_{\alpha_a}\}$, associate a vertex to the $i_a$-th output of 
channel $a$. Repeat the above construction for each of these vertices.  
\end{itemize}
In Fig.~\ref{fig:fnode} we show a small example of such a network.
We denote by $\Tree(i)$ the random (sub)network rooted at $i$, by
$\Tree(i,\ell)$ its first $\ell$ generations (starting from $i$),
and by $\uX_{i,\ell}$ the received colors, $\ell$ generations above
$i$ ( $\uX_{0,\ell}\equiv\uX_{\ell}$).

The network is used to communicate. A color $x_0\in\{1,\dots,q\}$
is chosen at the root with probability $\varphi_0(x_0)$ and
broadcast through  the $l_0$ channels connected to the root itself.
Each of the first generation vertices receives a corrupted 
version $x_i\in\{1,\dots,q\}$ of this color, with joint distribution
\begin{eqnarray}
\prod_{a=1}^{l_0}\pi^{(\alpha)}(x_{a,1},\dots,x_{a,i_a}|x_0)\, ,
\end{eqnarray}
where $(a,r)$ denotes the $r$-th output vertex of the $a$-th channel.
In other words distinct channels act independently. 
The same transmission process is repeated at the first generation and so 
forth, through the entire network. 
The problem is to reconstruct the transmitted color from the output 
at generation $\ell$, denoted as $\ux_{\ell}$.
 
Analogously to the case investigated in the previous sections,
we say that the reconstruction problem is solvable if the 
conditional mutual information
$I(X_0;\uX_{\ell}|\Tree)$ does not vanish as $\ell\to\infty$.
Equivalently, the problem is solvable if there is a reconstruction
procedure which succeeds with probability strictly larger than 
$\max_{x_0}\varphi_0(x_0)$ in the $\ell\to\infty$ limit. In these definitions
we assume the network structure to be known at the 
receiver (this is why we consider a  mutual information which is {\em conditional} to this structure). 

Unlike for $q$-ary reversible channels, the distribution of the 
color $X_i$ received at vertex $i$, is not uniform and depends upon the 
vertex. We shall denote it by $\varphi_i(x_i)\equiv \prob[X_i=x_i|\Tree]$.
In fact $\varphi_i(\, \cdot\,)$ can be determined recursively:
if $j$ is the parent of $i$, and $i_1$, $i_{k-1}$ the other vertices
that share this parent, then
\begin{eqnarray}
\varphi_i(x_i) = \sum_{x_j}\sum_{x_{i_1},\dots,x_{i_{k-1}}}
\pi^{(\alpha)}(x_i,x_{i_1},\dots,x_{i_{k-1}}|x_j)\,\varphi_j(x_j)\, .
\end{eqnarray}

Let us consider the reconstruction problem.
The conditional distribution of the transmitted color given the observation
at generation $\ell$, $\prob[X_0=x|\uX_{\ell}=\ux_{\ell}]$ can be computed
recursively proceeding from the leaves downwards to the root
as in Sec.~\ref{se:merging}.
In order to simplify the analysis, it is convenient to `factor out'
the {\em a priori} information $\varphi_i(x_i)$, and define
\begin{eqnarray}
\eta_{i,\ell}(x) = \prob_{i}[X_i=x|\uX_{i,\ell}]\, .
\end{eqnarray}
where $\prob_{i}$ denotes probability with respect to a modified 
process in which the boundary $ \uX_{i,\ell}$ is obtained from a broadcast
starting from  $X_i$  chosen uniformly at random in 
$\{1,\dots,q\}$. 
Of course we have
\begin{eqnarray}
 \prob[X_i=x|\uX_{i,\ell}] = \frac{1}{z_i}\, \varphi_i(x)\, 
\eta_{i,\ell}(x)\, ,
\end{eqnarray}
where $z_i \equiv \sum_x \varphi_i(x)\eta_{i,\ell}(x)$ ensures the 
correct normalization.
Notice that $\eta_{i,\ell}(x)$ depends uniquely on the portion 
of the tree above $i$, more precisely on $\Tree(i,\ell)$ and $\uX_{i,\ell}$.

With a slight abuse of notation, let us denote by $\pi^{(a)}=\pi^{\alpha(a)}$, 
$a \in\{ 1,\dots, l_i\}$ be the channels whose input is $x_i$,
and by $j_1,\dots,j_{k(a)}$ the corresponding output vertices.
It is easy to derive the following recursion
which generalizes Eq.~(\ref{eq:Site})
\begin{eqnarray}
\eta_{i,\ell+1}(x) = \frac{1}{z(\{\eta_{j,\ell}\},\{\pi^{(a)}\})}
\,\prod_{a=1}^{l_i}\sum_{x_{1},\dots x_{k(a)}}
\pi^{(a)}(x_{1},\dots x_{k(a)}|x)\;
\eta_{j_1,\ell}(x_1)\cdots\eta_{j_{k(a)},\ell}(x_{k(a)})\, ,
\end{eqnarray}
where $\ell$ is the distance from the leaves.
The constant $z(\{\eta_{j,\ell}\},\{\pi^{(a)}\})$ is defined
by the constraint $\sum_{x}\eta_{i,\ell+1}(x) =1$. 
We shall denote the above mapping synthetically by writing
$\eta_{i,\ell+1} = \F(\{\eta_{j,\ell}\},\{\pi^{(a)}\})$.

One can define two types of probability distributions 
associated to $\eta_{i,\ell}$. We first assume that  the
tree $\Tree$ is given, and  
consider the distribution of $\eta_{i,\ell}$ conditional to $X_i=x$
and $\Tree$. This will be denoted by $Q^{(i,\ell)}_x(\eta)$.
Arguing as in the case of regular trees, one derives the recursion
\begin{eqnarray}
Q^{(i,\ell+1)}_x(\eta) =\sum_{\{x_j\}} \prod_{a=1}^{l_i}
\pi^{(a)}(x_{j_1},\dots x_{j_{k(a)}}|x)\;
\int\;\delta [\eta- \F(\{\eta_{j}\},\{\pi^{(a)}\})]\;
\prod_j \de Q^{(j,\ell)}_{x_j}(\eta_j)\, .\label{eq:GeneralIteration}
\end{eqnarray}

Next, we consider the distribution of $\eta_{i,\ell}$
{\em unconditional} of $\Tree$, which we denote by $Q^{(\ell)}_x(\eta)$.
This can also be regarded as the expectation of the previous
distribution: $Q^{(\ell)}_x(\eta) = \E_{\Tree} Q^{(i,\ell)}_x(\eta)$.
Notice that  $Q^{(i,\ell)}_{x}(\eta)$ depends on 
$\Tree$ only through the first $\ell$ generations of the subtree rooted at $i$
$\Tree(i,\ell)$. Since the structures of distinct subtrees are
independent, if we average Eq.~(\ref{eq:GeneralIteration})
with respect to $\Tree$, the averages factorize, yielding
a recursion equation for $Q^{(\ell)}_x(\eta)$:
\begin{eqnarray}
Q^{(\ell+1)}_x(\eta) = \E\sum_{\{x_j\}} \prod_{a=1}^{l_i}
\pi^{(a)}(x_{j_1},\dots x_{j_{k(a)}}|x)\;
\int\;\delta [\eta- \F(\{\eta_{j}\},\{\pi^{(a)}\})]\;
\prod_j \de Q^{(\ell)}_{x_j}(\eta_j)\, .\label{eq:GeneralIterationSimpl}
\end{eqnarray}
Here $\E$ denotes expectation with respect to the degree $l_i$ and the
channel types. The last expression is particularly convenient for 
numerical simulations and it is not more complex than the iteration 
(\ref{eq:Iteration}) studied in the previous Sections.

We can also consider the distributions unconditional to
the transmitted color: $\Q^{(i,\ell)}(\eta)$ and $\Q^{(\ell)}(\eta)$.
The same relation as for regular trees
hold in this case $Q^{(i,\ell)}_x(\eta) = q\eta(x) \Q^{(i,\ell)}(\eta)$
and $Q^{(\ell)}_x(\eta) = q\eta(x) \Q^{(\ell)}(\eta)$. It is easy to
derive the corresponding recursions. We just
write down the equation for the last (non-random) distribution
\begin{eqnarray}
\Q^{(\ell+1)}(\eta) = \E\;
\int\;\frac{z(\{\eta_j\}\{\pi^{(a)}\})}{z(\{\eo\}\{\pi^{(a)}\})}\;\;
\delta [\eta- \F(\{\eta_{j}\},\{\pi^{(a)}\})]\;
\prod_j \de \Q^{(\ell)}(\eta_j)\, .\label{eq:SymmetricGeneral}
\end{eqnarray}

In order to discuss the correspondence with the dynamical glass transition 
in this more general setting, it is necessary to distinguish two cases. 
In the simplest one, the RS cavity equations for the associated statistical
mechanics model admit the solution $\eta_{i\to j}(x) =\eo(x)$
for any directed link $i\to j$ in the graph. Under this hypothesis
it is not hard to show  that Eq.~(\ref{eq:SymmetricGeneral}) is equivalent 
(in the same sense discussed in Sec.~\ref{sec:Recursion}) to the
$m=1$ 1RSB equation~\footnote{The expert will perhaps be surprised
by this remark since it is usually said that the order parameter for
such systems is a `measure over the space of distributions'. 
However
it turns out that, for $m=1$, the expectation of this measure
satisfies an equation which is Eq.~(\ref{eq:SymmetricGeneral}).}

In the general case (i.e. if  $\eta_{i\to j}(x) =\eo(x)$ does not solve the 
RS equations), the dynamical glass transition still corresponds to 
the extremality of the free boundary measure on an infinite tree. 
The last problem, however, cannot be formulated in the same framework as 
described here. One can still write an equation of the form 
(\ref{eq:GeneralIteration}), conditioned to the graph structure, as is usually
done in statistical physics. But the average over the graph structure can only be 
performed conditioning upon the value of $\eta_{i\to j}(\,\cdot\,)$
in the RS solution~\cite{ReconstrNext}, and so the relation between the spin glass problem and the
reconstruction problem unconditioned to the structure of the tree, is not as simple as before. We
shall not enter these details here.  
%
%
\section{Conclusion and open problems}
\label{sec:Conclusion}

The coincidence between the reconstruction threshold and the dynamical glass
transition explored in this paper is interesting from several points
of view.

First of all, it provides some  perspective for putting on a firmer
basis the theory of glassy systems on locally tree-like graphs 
developed in the last few years. 
As a concrete example, notice that the population dynamics algorithm
defined in Section \ref{sec:Ferro} presents two important 
advantages with respect to the procedure adopted by statistical physicists. 
First, in spin glass theory the usual
prescription is to look for a solution of the fixed point
equation (\ref{eq:1RSB_fixed}). This poses the problem of the initial
condition: even if a given initial condition yields a trivial fixed point,
this may not be the case for all the initial conditions. In the present 
formulation the iteration is initialized with a very specific initial 
condition, and one is guaranteed that, if it converges to a trivial fixed 
point, no non-trivial fixed point exists.  
Second, simulating Eq.~(\ref{eq:1RSB_fixed}) 
requires a `reweighting' of the sample which is usually the trickier
part of the calculation. No reweighting is needed in the new approach:
Eq.~(\ref{eq:Iteration}) can be handled easily.

Also, it provides some indication on the correctness of simple 
analytic and algorithmic approaches to these systems, 
such as the replica symmetric cavity method, or the belief 
propagation algorithm. While it has long been known that their 
correctness should be related to a fast correlation decay in a
model on a tree, a precise criterion has never been 
formulated
\footnote{A frequently used sufficient  criterion is the uniqueness of
the Gibbs state on the infinite tree (see \cite{BanGam} and references 
therein). As it emerges from our discussion, this criterion is often much stronger 
 than needed. It is also interesting to recall that 
Tatikonda and Jordan \cite{Tatikonda} first connected the convergence 
properties of belief propagation to the extremality of the free boundary Gibbs 
on a properly defined tree.}. 
Ous work suggests that the extremality of  the associated Gibbs measure
on a tree provides such a criterion.

More broadly, it illustrate some subtleties of the physics of
the glass transition. It is well known that, in a 1RSB
glass transition, point to point correlations (static scattering factors)
do not present any diverging correlation length. This paper
shows that such a length can be derived quite generally from
point-to-set correlations. Indeed the definition considered here is 
essentially equivalent to the one of Ref.~\cite{BoBi04}.
It was shown in Ref.~\cite{MoSe05} that this length scale divergence 
implies a lower bound on the time scale divergence.

Finally, statistical physics ideas can inspire new results
on the original reconstruction problem. The most interesting such
idea is, in our view, the complexity functional introduced in 
Section \ref{se:varprinc}. Apart from being conceptually innovative
with respect to classical techniques, it seems to provide by
far the best rigorous quantitative estimates of the reconstruction 
threshold, cf. Proposition \ref{propo:VariationalAnti}. 
This is well illustrated by the results in Table
 \ref{TableAnti}. It will certainly be interesting 
to find a concrete interpretation of this object in terms of
the original reconstruction problem. Also, the values of the channel parameter
at which the asymptotic complexity $\Sigma(\Q^{(\infty)})$ vanish have
a particularly important role in statistical mechanics, but did not
find any role here.

There are several results that we have not been able to  prove rigorously.
We already formulated one such results as Conjecture \ref{conj:Sigma},
and proved it for a particular class of channel models as Proposition
\ref{propo:VariationalAnti}. Another interesting fact which has emerged from our
numerical simulations is the coincidence of the KS and reconstruction 
thresholds when the number of colors is small.
\begin{conj}\label{conj:KS}
Consider the reconstruction problem for the $k$-ary tree and the 
ferromagnetic Potts channel ($q$-ary symmetric channel) with $q\le 4$,
or the antiferromagnetic Potts channel with $q\le 3$
Then, there exists a $k_{\rm max}\ge 30$ such that, if $k<k_{\rm max}$
the reconstruction threshold coincides with the Kesten-Stigum threshold.
\end{conj}
A stronger version of this conjecture would be to require the thesis 
to be valid for {\em all} values of $k$ (i.e. to state 
that $k_{\rm max} = \infty$). Although we didn't find any $k$ 
contradicting this stronger version, this might of course be due to
the limitation on the values of $k$ that we can  treat numerically.

Finally, let us single out the case of completely antiferromagnetic
($\varepsilon=1$) Potts channels:
\begin{conj}\label{conj:Col}
Let $k_*(q)$ be the maximum value of $k$ such that the free boundary 
Gibbs measure for uniformly random proper colorings on the infinite $k$-ary
tree is extremal. Then  $k_*(3) = 5$, $k_*(4)=8$, $k_*(5)=13$,
$k_*(6)=17$.
\end{conj}
%
%
\section*{Acknowledgments} 
This work has been carried out mostly during our stay
at the MSRI on the occasion of the program on ``Probability, Algorithms, and
Statistical Physics''. We have benefited from numerous discussions,
suggestions and comments from James Martin, Elchanan Mossel and Yuval Peres; 
it is a great pleasure to thank them for these stimulating exchanges. 
Useful discussions with Olivier Rivoire and Guilhem Semerjian
are also gratefully acknowledged.

%
%
\appendix 

\section{Variational principle for frustrated kernels}
\label{app:Variational}

This appendix is devoted to the proof of Lemma \ref{lemma:Variational}.
It is convenient to introduce some notations. If
$A(x,y)$, $x,y\in\{1,\dots,q\}$ is a symmetric matrix and 
$\eta_1(x)$, $\eta_2(x)$,  $x\in\{1,\dots,q\}$ are two vectors,
we shall write $A\eta_1(x)\equiv\sum_y A(x,y)\eta_1(y)$ and
$\eta_1 A\eta_2\equiv\sum_y A(x,y)\eta_1(x)\eta_2(y)$ .
Furthermore, if $\Q$ is a distribution over $\M$ we will
denote by $\T\Q$ the distribution obtained 
by using  Eq.~(\ref{eq:1RSB_fixed}): $\T\Q$ is the left hand side
of Eq.~(\ref{eq:1RSB_fixed}) when in the right hand side  $\Q^*$
has been substituted by $\Q$.
 Finally, given 
$\eta_1,\eta_2\in\M$, we let 

\begin{eqnarray}
\Delta(\eta_1,\eta_2) \equiv \left(\frac{\eta_1\pi\eta_2}{\eo\pi\eo}\right)
\log\left(\frac{\eta_1\pi\eta_2}{\eo\pi\eo}\right)\, .
\end{eqnarray}

We also write $\eta\ed \Q$ when $\eta$ has distribution $\Q$.
We first derive two simple lemmas.
\begin{lemma}\label{remark1}
Let $\Q^*$ and $\Q$ be two consistent distributions over $\M$ and 
$\Sigma^*(t) \equiv\Sigma((1-t)\Q^*+t\Q)$. Then
\begin{eqnarray}
-\frac{1}{(k+1)}\left.\frac{\de\Sigma^*}{\de t}\right|_0 = 
\E\left\{\Delta(\nu,\eta_2')-\Delta(\nu,\eta_2)-\Delta(\eta_1,\eta_2')
+\Delta(\eta_1,\eta_2)\right\}\, ,
\label{eq:der_result}
\end{eqnarray}
where the expectation is taken with respect to the independent random variables
$\eta_1,\eta_2 \ed \Q^*$, $\eta_2'\ed \T\Q^*$ and $\nu\ed \Q$. 
\end{lemma}
\prooft
Elementary calculus yields
\begin{eqnarray}
-\frac{1}{(k+1)}\left.\frac{\de\Sigma^*}{\de t}\right|_0 =\psi(1)-\psi(0)\, ,
\label{eq:derivee}
\end{eqnarray}
where
\begin{eqnarray}
\psi(t) & \equiv &-\E\left\{\left(\frac{\nu^t\pi\eta}{\eo\pi\eo}\right)
\log\!\left(\frac{\nu^t\pi\eta}{\eo\pi\eo}\right)\right\}+\\
&& +\E\left\{ \left(\frac{\sum_x\pi\nu^t(x)\prod_{i=1}^k
\pi\eta_i(x)}{\sum_x\prod_{i=0}^k \pi\eo(x)}\right)\log\!
\left(\frac{\sum_x\pi\nu^t(x)\prod_{i=1}^k
\pi\eta_i(x)}{\sum_x\prod_{i=0}^k \pi\eo(x)}\right)\right\}\, .\nonumber
\end{eqnarray}
Here $\eta,\eta_1,\dots,\eta_k\ed \Q^*$ and $\nu^t\ed (1-t)\Q^*+t\Q$ 
are independent random variables. The first term, when integrated on $t$,
gives the contribution $\E\left\{\Delta(\eta_1,\eta_2)-\Delta(\nu,\eta_2)\right\}$
to (\ref{eq:der_result}).

As for the second term, 
observing that 
$\prod_{i=1}^k \pi\eta_i(x) = z(\{\eta_i\})\, \F(\eta_1,\dots,\eta_k)$,
it  can be rewritten as
\begin{eqnarray}
q^{k-1}\E\left\{z(\{\eta_i\}) \left[ \frac{\nu^t\pi\F(\eta_1\dots\eta_k)}
{\eo\pi\eo}\right]\log\left[ \frac{\nu^t\pi\F(\eta_1\dots\eta_k)}
{\eo\pi\eo}\right] \right\}+\nonumber\\
q^{k-1}\E\left\{z(\{\eta_i\}) \left[ \frac{\nu^t\pi\F(\eta_1\dots\eta_k)}
{\eo\pi\eo}\right]\log\left[ q^{k-1}z(\{\eta_i\})\right] \right\}\, .
\end{eqnarray}
Since $\E\nu^t(x) =\eo(x)$, the second term is $t$-independent 
 and does not contribute to (\ref{eq:derivee}). The first term is  equal
to $\E\Delta(\nu^t,\eta'_2)$ where $\eta'_2\ed\T\Q^*$.
\endproof

\begin{lemma}\label{remark2}
Let $\eta_1,\eta_2\in\M$ and define $\delta\eta_i(x) = \eta_i(x)-\eo(x)$.
If $\pi(x,y)=\pi_*-\pih(x,y)$ is a frustrated kernel, then 
\begin{eqnarray}
|\delta\eta_1\pih\delta\eta_2|\le \eo\pi\eo\, .\label{eq:remark2}
\end{eqnarray}
\end{lemma}
\prooft
Since $\pih$ is positive definite, $\phi\pih\psi\equiv\sum_{x,y}
\pih(x,y)\phi(x)\psi(y)$ is a well defined scalar product.
Cauchy-Schwarz inequality implies
\begin{eqnarray}
|\delta\eta_1\pih\delta\eta_2|\le\sqrt{(\delta\eta_1\pih\delta\eta_1)
(\delta\eta_2\pih\delta\eta_2)}\le \max\left\{(\delta\eta_1\pih\delta\eta_1),
(\delta\eta_2\pih\delta\eta_2)\right\}\,. 
\end{eqnarray}
Therefore it is sufficient to prove Eq.~(\ref{eq:remark2}) 
for $\delta\eta_1=\delta\eta_2=\delta\eta$. Let $\eta(x) = \eo(x)+
\delta\eta(x)$. Since $\pi(x,y),\eta(x)\ge 0$, 
and $\sum_x\delta\eta(x) = 0$, we have
\begin{eqnarray}
0\le\eta\pi\eta = \eo\pi\eo+\delta\eta\pi\delta\eta=\eo\pi\eo-
\delta\eta\pih\delta\eta\, .
\end{eqnarray}
\endproof

We can now turn to the proof of Lemma \ref{lemma:Variational}. In the 
following, given $\eta\in\M$, we define $\delta\eta(x)\equiv
\eta(x)-\eo(x)$. Obviously we have
\begin{eqnarray}
\Delta(\eta_1,\eta_2) = \left(1-\frac{\delta\eta_1\pih\delta\eta_2}{\eo\pi\eo}\right)
\log\!\left(1-\frac{\delta\eta_1\pih\delta\eta_2}{\eo\pi\eo}\right)\, .
\end{eqnarray}
Because of Lemma \ref{remark2}  we can expand this expression in an 
absolutely convergent series
\begin{eqnarray}
\Delta(\eta_1,\eta_2) = -\frac{\delta\eta_1\pih\delta\eta_2}{\eo\pi\eo}
+\sum_{n=2}^{\infty} C_n\, 
\left(\frac{\delta\eta_1\pih\delta\eta_2}{\eo\pi\eo}\right)^n\, ,
\end{eqnarray}
where $C_n\equiv 1/n(n-1)>0$. If $\eta_1\ed \Q_1$ and $\eta_2\ed\Q_2$ are 
independent random variables with consistent distributions, we get
\begin{eqnarray}
\E\Delta(\eta_1,\eta_2) = \sum_{n=2}^{\infty} C_n\,q^n
\phi_1^{(n)}\pih^{\otimes n}\phi_2^{(n)}\, ,
\label{eq:Expansion}
\end{eqnarray}
where $\pih^{\otimes n}(x_1\dots x_n;y_1\dots y_n) \equiv \pih(x_1,y_1)
\cdots \pih(x_n,y_n)$ is the $n$-fold
tensor product of $\pih$, $\phi_a^{(n)}(x_1\dots x_n)
\equiv \E\{\delta\eta_a(x_1)\cdots \delta\eta_a(x_n)\}$ 
are the moments of the distribution $\Q_a$, and
\begin{eqnarray}
\phi_1^{(n)}\pih^{\otimes n}\phi_2^{(n)} \equiv \sum_{x_1\dots x_n}
\sum_{x_1\dots x_n} \pih^{\otimes n}(x_1\dots x_n;y_1\dots y_n)
\phi_1^{(n)}(x_1\dots x_n)\phi_2^{(n)}(x_1\dots x_n)\, .
\end{eqnarray}

Now consider Remark \ref{remark1} and take $\Q = \T\Q^*$. We get 
\begin{eqnarray}
-\frac{1}{(k+1)}\left.\frac{\de\Sigma^*}{\de t}\right|_0 = 
\E\left\{\Delta(\eta_1',\eta_2')-\Delta(\eta_1',\eta_2)-\Delta(\eta_1,\eta_2')
+\Delta(\eta_1,\eta_2)\right\}\, ,
\end{eqnarray}
where $\eta_1,\eta_2\ed\Q^*$ and $\eta_1',\eta_2'\ed\T\Q^*$. Applying
Eq.~(\ref{eq:Expansion}) we get 
\begin{eqnarray}
-\frac{1}{(k+1)}\left.\frac{\de\Sigma^*}{\de t}\right|_0 = 
\sum_{n=2}^{\infty} C_n\,q^n
(\phi_{\T}^{(n)}-\phi^{(n)})\pih^{\otimes n}
(\phi_{\T}^{(n)}-\phi^{(n)})\, ,\label{eq:LastProofVariational}
\end{eqnarray}
where $\phi^{(n)}$ and $\phi^{(n)}_{\T}$ denote the moments (respectively)
of $\Q^*$ and $\T\Q^*$. Since $\pih$ is positive definite,
 $\pih^{\otimes n}$ is positive definite as well and therefore the
right hand side is a sum of non-negative terms. In order for this right hand side
to vanish, each of the terms must vanish, which implies
$\phi_{\T}^{(n)}=\phi^{(n)}$ for each $n$. But, since $\Q^*$ and 
$\T\Q^*$ have bounded support, this implies $\Q^* = \T\Q^*$, which
is false by hypothesis. Therefore the right hand side of 
Eq.~(\ref{eq:LastProofVariational}) is strictly positive and 
$\left.\frac{\de\Sigma^*}{\de t}\right|_0 <0$ as desired. \endproof
%
%
\bibliographystyle{alpha}

\newcommand{\etalchar}[1]{$^{#1}$}


\end{document}